\documentclass[12pt,tightenlines,eqsecnum,floats,aps,amsmath,amssymb,nofootinbib,prd]{revtex4}

\usepackage{setspace}
\usepackage{amsmath,amssymb,amsfonts,amsthm,mathrsfs}
\usepackage{graphicx}
\usepackage{enumerate} 

\usepackage[pdftex]{hyperref}

\def\be{\begin{equation}}
\def\ee{\end{equation}}
\def\ba{\begin{eqnarray}}
\def\ea{\end{eqnarray}}
\def\bi{\begin{itemize}}
\def\ei{\end{itemize}}

\def\rationals{\mathbb{Q}}

\def\integers{\mathbb{Z}}
\def\reals{\mathbb{R}}
\def\complex{\mathbb{C}}

\def\G{\mathcal{G}}
\def\tr{\text{Tr}}

\def\bra{\langle}
\def\ket{\rangle}

\def\id{\text{Id}}

\def\w{\omega}

\def\t{\tau}

\def\lqg{\text{\tiny LQG}}

\def\Hks{\mathcal{H}_{\text{KS}}}
\def\P{\mathcal{P}}
\def\E{\mathcal{E}}
\def\A{\mathcal{A}}
\def\Abar{\bar{\mathcal{A}}}
\def\hba{\mathcal{HBA}}
\def\hbabar{\overline{\mathcal{HBA}}}
\def\Ba{\mathcal{BA}}
\def\Babar{\overline{\mathcal{BA}}}
\def\ha{\mathcal{HA}}
\def\habar{\overline{\mathcal{HA}}}

\def\Ah{\bar{\mathcal{A}}_\text{H}}
\def\Ab{\bar{\mathcal{A}}_\text{B}}

\def\Lh{\mathcal{L}_\text{H}}
\def\Lb{\mathcal{L}_\text{B}}
\def\L{\mathcal{L}}
\def\spec{\Delta}
\def\muks{\mu_{\text{KS}}}

\def\cyl{\text{Cyl}}
\def\hom{\text{Hom}}

\def\p{\tilde{p}}
\def\Eb{E}
\def\pol{\text{Pol}}
\def\gen{\text{Gen}}
\def\Vder{V_{\text{deriv}}}
\def\Xh{X^{\text{H}}}
\def\Xb{X^{\text{B}}}
\def\et{\tilde{e}}
\def\spech{\Delta_\text{H}}
\def\specb{\Delta_\text{B}}
\def\s{\tilde{s}}
\begin{document}
\title{The Koslowski-Sahlmann representation:
Quantum Configuration Space
}
\author{Miguel Campiglia} \email{miguel@rri.res.in}
\author{Madhavan Varadarajan}\email{madhavan@rri.res.in}
 \affiliation{Raman Research Institute \\Bangalore-560 080, India}

\begin{abstract}

The Koslowski-Sahlmann (KS) representation is a
generalization of the  representation underlying the discrete spatial geometry of  Loop Quantum Gravity (LQG), 
to accommodate states labelled by smooth spatial 
geometries. As shown recently,  
the KS representation supports, in addition
to the action of the  holonomy and flux operators,  
the action of operators which are the quantum counterparts of certain
connection dependent functions known as ``background exponentials''. 

Here we show that the
KS representation displays the following  properties which are  the exact counterparts of 
LQG ones: 
(i) the abelian $*$ algebra of $SU(2)$ holonomies and `$U(1)$'
background exponentials can be completed to a $C^*$ algebra
(ii) the space of semianalytic $SU(2)$ connections is topologically dense in the spectrum of this algebra 
(iii) there exists a measure on this spectrum for which  the KS
Hilbert space is realised as the space of square integrable functions on the spectrum 
(iv) the spectrum admits a characterization as a projective limit of  finite numbers of copies of 
$SU(2)$ and $U(1)$ 
(v) the algebra underlying the KS representation is 
constructed from cylindrical functions and their derivations in exactly the same way as the LQG (holonomy-flux) algebra 
except that the KS cylindrical
functions depend on the holonomies  {\em and} the
background exponentials, this extra dependence being responsible for the differences between the KS and LQG algebras.

While these results are obtained for  compact spaces, they  are expected to 
be of use
for the construction of the KS representation in the asymptotically flat case.

\end{abstract}
\maketitle

\section{Introduction} \label{sec1}

Loop Quantum Gravity (LQG) is an attempt at canonical quantization of a  
classical Hamiltonian description of gravity in terms of an $SU(2)$ 
connection and its conjugate electric field on a Cauchy slice. 
The electric field
plays the role of a triad and thereby endows the slice with a spatial geometry.
One of the key results of LQG is that the corresponding quantum geometry
associated with LQG states has a fundamental discreteness.
The smooth classical geometry of space is then expected to arise through 
a suitably coarse grained view of this discrete geometry \cite{weave}.

One may enquire as to whether it is possible to describe 
the effective smoothness of classical spatial
geometry directly at the quantum level without explicit recourse to 
any coarse graining. Koslowski answered this question affirmatively by 
slightly modifying the standard LQG representation \cite{kos}. He did this through an  
assignation of   an additional 
smooth triad field label to every kinematic LQG state in conjunction with a modification 
of the action of flux operators so as to make them sensitive to the additional label. 
As demonstrated in detail by Sahlmann \cite{hs}, 
the area and  volume operators then
acquire, in addition to the standard LQG type discrete 
contributions, 
a `smooth' contribution determined by this additional label. 

As explained in detail elsewhere \cite{miguelme}, our interest in the 
Koslowski-Sahlmann (KS) representation arises from the possibility of
using it to explore asymptotic flatness in canonical quantum gravity.
As a precursor to such an exploration, it is of interest to obtain a
detailed  understanding of the KS representation for the simpler 
case of compact spatial topology. Accordingly, building on the work
of Koslowski and Sahlmann, we initiated a study of the KS representation
in References \cite{me,miguelme}. 

In Reference \cite{me}, it was shown that in addition to the 
holonomy and flux operators of LQG, the KS representation also supports 
the action of the quantum correspondents of 
certain classical functions of the connection  called ``background
exponentials''. Each such exponential $\beta_{\bar E}(A)$ is labelled by a background electric field ${\bar E}^a_i$
and defined as $\beta_{\bar E}(A):= \exp(i\int_{\Sigma} {\bar E}^a_iA_a^i)$. 
Building on this work and that of Sahlmann, 
in Reference \cite{miguelme}, we studied
the imposition of gauge and diffeomorphism invariance in the KS representation.

In this work we further study the KS representation with a view to 
providing structural characterizations 
similar to those developed for  LQG.
We refer here to the  beautiful developments in the field, mainly in the nineties, 
which provided a characterization of the
LQG Hilbert space as that of  square integrable functions on a quantum 
configuration space of 
`generalized' connections, 
the square integrability being defined
 with respect to  a 
suitable `Ashtekar-Lewandowski' measure on this space 
\cite{aaisham,aajurekhoop,alm2t,vel1st,velhinho,ttbook}.
Moreover, this quantum 
configuration space can be viewed as a projective limit space 
\cite{josedon,aajurekproj,alm2t,velhinho,vel1st,ttbook}.
In this work we prove that exact counterparts of these characterizations
exist for the KS representation. The layout of our paper, including a detailed
description of our results, is as follows.

The results in the paper depend on the use of structures which relate to the algebraic properties of 
holonomies and background exponentials. The structures associated with 
holonomies are semianalytic edges, piecewise semianalytic curves, the groupoid of paths and its subgroupoids.
These holonomy related structures are used to show the classic LQG results mentioned above (see for example
\cite{velhinho,ttbook}). 
The structures associated with  background exponentials are the vector space 
of 
semianalytic $SU(2)$ electric fields, its abelian group structure under addition,
and the  subgroups of this group which are 
generated by  sets of {\em rationally independent} semianalytic electric fields. 
These sets of  rationally independent semianalytic electric fields are the background exponential 
related counterparts of the 
sets  of {\em independent} edges, 
the latter serving as sets of independent 
`probes' of the space of connections in the LQG context \cite{aajurekhoop,velhinho,ttbook}.
Section \ref{sec2} serves  to review the above holonomy related structures (leaning heavily on the exposition of 
\cite{velhinho,vel2,ttbook})  as well as to define the background exponential related ones.
Section \ref{sec2} also establishes our notation for the rest of the paper.

Most of our results depend on the validity  of
a key `Master Lemma'. This  Lemma  states that, given a set of independent probes and a corresponding set of 
elements in $SU(2)$ (one for each independent edge) and $U(1)$ (one for each rationally independent electric field),
there exists a semianalytic connection such that the evaluation of the relevant set of holonomies and 
background exponentials on this connection reproduces the given set of group elements to arbitrary accuracy.
In section \ref{sec3}, we provide a precise statement of this Lemma and describe the idea behind its proof. The 
proof itself is  technically involved and relegated to an appendix.

Section \ref{sec4} is devoted to the derivation of $C^*$ algebraic results for the KS representation.
First, we show that the abelian 
Poisson bracket algebra of holonomies and background exponentials, 
$\hba$,
 can be completed to a $C^*$ algebra, $\hbabar$. From general $C^*$ algebraic arguments \cite{rendall}, one  concludes that 
the classical configuration space of connections is densely embedded in
the  Gel'fand  spectrum $\spec$ of 
$\hbabar$, so that the spectrum may be thought of as a space of
`generalised' connections. In order to understand elements of this space
better, we show that every element of the spectrum 
is in correspondence
with a pair of homomorphisms, one homomorphism from the path groupoid to 
$SU(2)$ and
the other  from the abelian group of electric fields to $U(1)$. The first homomorphism
corresponds to the algebraic structure provided by the holonomies and the second to that 
provided by the background exponentials.

Next we turn our attention to the definition of a measure on the spectrum 
which allows the identification of the spectrum as the quantum configuration
space for the KS representation.
We show that the KS `vacuum expectation' value defines a positive
linear function on $\hbabar$. Standard theorems then imply that
this function defines a measure $d\muks$ on the  spectrum 
$\spec$
and that the KS Hilbert space
is isomorphic to the space $L^2( \spec, d\muks) $.
Next, we  consider the electric flux operators.
These operators map the finite span of  KS spinnets
into itself. We define the action of these operators on 
$L^2( \spec, d\muks) $
through the identification of KS spin 
network states with appropriate cylindrical functions on the spectrum. The compatibility of the measure $d\muks$
with the adjointness properties of the flux operators so defined, follows immediately from the fact that these
adjointness properties are implemented in the KS representation.

Section \ref{sec5} is devoted to the projective limit characterization of the quantum configuration space.
We show that the spectrum $\spec$
is homeomorphic to an appropriate 
projective limit space $\Abar$ 
whose fundamental building blocks are products of 
finite copies of $SU(2)$ and $U(1)$. Once again, the $U(1)$ copies capture the structure provided by the 
background exponentials whereas the $SU(2)$ copies correspond to, as in LQG, the structure provided by the 
 holonomies.
Following Velhinho \cite{velhinho},
we show this through the  
identification of the $C^*$- algebraic and the projective limit
 notions of cylindrical functions 
together with an appropriate application of the 
Stone-Weierstrass theorem. In this context, as we shall show, 
the $C^*$ algebraic notion of cylindrical functions corresponds to polynomials in the holonomies {\em and} background 
exponentials and their projective analogs to the same polynomials with holonomies replaced by $SU(2)$ elements
and background exponentials by $U(1)$ elements. Next, we show that the Haar measures on these building blocks define a consistent family of cylindrical measures on $\Abar \equiv \spec$ and that this family derives from the KS measure $d \muks$ on $\spec$. We use this characterization of $d\muks$ to show that, similar to LQG \cite{josedon}, $\A$ lies in a zero measure set within $\Abar$.

Section \ref{sec6} focuses on the analysis of the algebraic structure underlying the KS representation.
Our motivation for such an analysis stems from recent work by Stottmeister and Thiemann (ST) \cite{ttks} 
in which they point out that the KS representation is not a consistent representation of the standard
holonomy-flux algebra of LQG. ST point out  that any representation of the holonomy-flux algebra must satisfy
an infinite number of identities involving flux operators and their commutators. 
They provide an  explicit and beautiful example relating the 
double commutator of a triplet of  fluxes  to a single flux \cite{ttks} and they show
 that the KS representation does not 
satisfy this identity. Since the holonomies and fluxes are well defined operators in the KS representation, 
this `Stottmeister-Thiemann' obstruction 
calls into question the existence of a consistent algebraic structure underlying the 
KS representation. In section \ref{sec6} we explicitly construct exactly such a consistent algebraic structure. 

We proceed as follows.
Recall that the construction of the standard holonomy-fux algebra, including the precise non-commutativity of the 
fluxes, is  based on work of  Ashtekar, Corichi and Zapata (ACZ) \cite{acz}. The fluxes are functions on the 
phase space of gravity. ACZ studied the action of the Hamiltonian vector fields associated with these fluxes
on functions of connections constructed out of holonomies. This action is that of a derivation and
ACZ captured the non-commutativity of the flux operators in LQG through the non-commutativity of
these (classical) derivations. This in turn led to the construction of the holonomy-flux algebra in terms
of (cylindrical) functions of holonomies and their derivations.
Accordingly, 
we generalise the ACZ 
considerations to the KS case wherein the space of connection dependent functions
is built out of not only the holonomies but also the background exponentials.
The ACZ arguments so generalised indicate the identification of
the algebraic structure of the Poisson bracket between fluxes  with the
commutator of  
derivations on this (enlarged) space of (cylindrical) functions. 
This implies that similar to the considerations of Reference \cite{lost},
the KS counterpart of the holonomy flux algebra is generated by these 
functions, their derivations (which are 
obtained through the action of the flux Hamiltonian vector fields)  and multiple commutators
of these derivations. We call this algebra as
the holonomy--background exponential--flux algebra. We show, from 
the considerations of sections \ref{sec3} and \ref{sec4}, that
the KS representation  is  indeed  a representation of the 
holonomy--background exponential--flux algebra. 

This algebra is 
different from the usual holonomy-flux algebra by virtue of the extra structure provided by the background 
exponentials. In particular, the derivations corresponding to the flux Hamiltonian vector fields
have, so to speak, an extra set of $U(1)$ components in addition to the usual $SU(2)$ ones. 
It is this extra structure, directly traceable to the background exponential functions, which is responsible
for the evasion of the Stottmeister-Thiemann obstruction. In other words: (i) there {\em is} a consistent 
algebraic structure underlying the KS representation, namely the holonomy--background exponential--flux algebra, 
(ii) this structure is different from the standard
holonomy-flux algebra, and, (iii) this structure does not necessarily support the flux commutator identities
which are satisfied by representations of the holonomy-flux algebra; in particular, as we explicitly show,
it does not satisfy the triple flux commutator identity of \cite{ttks}.
This concludes our description of section \ref{sec6}.

Finally, 
Section \ref{sec7}  contains a brief discussion of our results as well as some remarks on the asymptotically flat case.

\section{Preliminaries} \label{sec2}

All differential geometric structures of interest will be based on the semianalytic, $C^k,k \gg 1$
category. The 
classical configuration space $\A$ is given by $su(2)$-valued one-forms $A_a = A^i_a \t_i  $ on a compact 
(without boundary) 3-manifold $\Sigma$.  $\t_i, i=1,2,3$ are $su(2)$ generators with $[\t_i,\t_j]=\epsilon_{i j k} \t_k$  and  $A_a \in \A$  represents  $SU(2)$ connections of a trivial bundle.   
The elementary configuration space functions are 
\begin{eqnarray}
h_{e }[A]_{C}^{\; D} &:= & (\P e^{\int_{e} A})_{C}^{\; D}  , \label{hol}\\
\beta_{E}[A] &:= &e^{i \int_\Sigma E \cdot A}  , \label{baexp}
\end{eqnarray}
where $C,D=1,2$, $h_{e }[A]_{C}^{\; D} \in SU(2)$ is the $j=1/2$ holonomy of the connection $A$ along an edge $e$ (see section \ref{sec2A}); $ E \cdot A\equiv E^a_i A_a^i$ with  $E^a=E^a_i \t^i$  a non-dynamical, unit density weight, smearing electric field.  $\beta_{E}[A]\in U(1)$ is referred to as a `background exponential' .


\subsection{Holonomy related structures} \label{sec2A}
The probes associated to holonomies are  \emph{paths}. A path $p$ is an equivalence class of oriented 
piecewise semianalytic curves on the manifold, where two curves are equivalent if they differ by orientation preserving reparametrizations and retracings.  $b(p)$ and $f(p)$ denote the beginning and end points of a path $p$. Two paths $p$ and $p'$ such that $f(p)=b(p')$ can be  composed to form a new path denoted by $p' p$. Thus $b(p'p)=b(p)$ and $f(p'p)=f(p')$. Under this composition rule, the set of all paths, $\P$, becomes a groupoid. 
An edge $e$ is a path $p$ that has a representative curve which is semianalytic, with the  image $\et$ of this representative curve being  a submanifold with boundary. Paths can always be written as finite compositions of edges. See \cite{velhinho,ttbook} for more precise definitions.

For a given  connection $A \in \A$, the holonomy along a path $p$, $h_p[A] \in SU(2)$, satisfies  $h_{p' p}[A]_{C}^{\; D}=h_{p'}[A]_{C}^{\; C'} h_{p}[A]_{C'}^{\; D}$, where $f(p)=b(p')$. Thus $A$ defines a  \emph{homomorphism} from $\P$ to $SU(2)$. The set of \emph{all} homomorphisms from $\P$ to $SU(2)$ is denoted by  $\hom(\P,SU(2))$, and corresponds to the space of generalized connections in LQG \cite{velhinho}.

A set of edges $e_1,\ldots,e_n$ is said to be independent if their intersections can only occur at their endpoints, i.e. if $\et_i \cap \et_j \subset \{b(e_i),b(e_j),f(e_i),f(e_j) \}$. We denote by $\gamma:=(e_1,\ldots,e_n)$ an ordered set of independent edges and by $\Lh$ the set of all such ordered sets of independent edges.  Given $\gamma,\gamma' \in \Lh$, we say that  $\gamma' \geq \gamma$ iff all edges of $\gamma$ can we written as composition of edges (or their inverses) of $\gamma'$. Equivalently, if $\P_\gamma$ denotes the  subgroupoid of $\P$ generated by edges of $\gamma$, then $\gamma' \geq \gamma$ iff $\P_\gamma$ is a subgrupoid of $\P_{\gamma'}$. It then follows that i)  $\gamma \geq \gamma$ and ii) $\gamma'' \geq \gamma', \gamma' \geq \gamma \implies \gamma'' \geq \gamma$ . Thus  `$\geq$' defines a {\em preorder} \cite{encytop} 
in $\Lh$. 

A preorder is weaker than a partial order in that it does not necessarily entail antisymmetry i.e.
$a\geq b, b\geq a$ does not necessarily imply $a=b$. For example,
$\gamma' \geq \gamma,\; \gamma \geq \gamma'$ does not imply that $\gamma = \gamma'$ since
the two relevant sets of edges may  differ in the ordering of their elements 
or by the substitution of an edge by its inverse.

Next, note that semianalyticity of the edges implies that given $ \gamma,\gamma' \in \Lh$ there always exists $\gamma''$ such that $\gamma'' \geq \gamma$ and $\gamma'' \geq \gamma'$ \cite{ttbook,velhinho}. Thus $(\Lh,\geq)$ is a directed set.\footnote{The label set $\Lh$  differs slightly from the one  used in Refs. \cite{velhinho,ttbook} where labels are given by subgroupoids $\P_\gamma$,  regardless of the choice of `generator' $\gamma$. One can nevertheless use $\Lh$ in the projective limit characterization of the quantum configuration space. See Appendix \ref{projlimapp} for  details.} 

\subsection{Background exponential related structures} \label{sec2B}
The probes associated to the  background exponentials are electric fields, and we denote by $\E$ the set of all (semianalytic) electric fields.  We will often see $\E$ as an Abelian group with composition law given by addition:  $(E, E') \to E+E' $. For a given  connection $A \in \A$, the background exponential function (\ref{baexp}) satisfies
 $\beta_{E'}[A] \beta_E[A]=\beta_{E'+E}[A]$ and thus defines an element in $\hom(\E,U(1))$  (the set of homomorphism from $\E$ to $U(1)$). A set of electric fields $E_1, \ldots, E_N$ will be said to be independent, if they are algebraically independent, i.e. if they are independent under linear combinations with integer coefficients:\footnote{It is easy to verify that (\ref{algind}) is equivalent to rational independence, i.e. the analogue of condition (\ref{algind}) with $q_I \in \rationals$.}
\be
\sum_{I=1}^N q_I E_{I}=0, \; q_I \in \integers \iff q_I =0 , I=1,\ldots,N .\label{algind}
\ee
We denote by $\Upsilon=(E_1, \ldots, E_N)$ an ordered set of independent electric fields. The set of all ordered  sets of independent electric fields is denoted by $\Lb$.  Let $\E_\Upsilon = \integers E_1 + \ldots +\integers E_N \subset \E$  denote the subgroup of $\E$ generated by $\Upsilon$. We then define $\Upsilon' \geq \Upsilon$  iff $\E_\Upsilon$ is a subgrup of $\E_{\Upsilon'}$, or  equivalently if the electric fields in $\Upsilon$ can be written as algebraic combinations of  those in $\Upsilon'$. As in the edge case, it follows that $\geq$ is a preorder relation.  

In appendix \ref{gensetapp} it is shown that given any finite set of electric fields (not necessarily independent), there always exists a finite set of algebraically independent electric fields that generates the original set. Applying this result to  the set $\Upsilon \cup \Upsilon'$ for given $\Upsilon,\Upsilon' \in \Lb$, we find $\Upsilon'' \in \Lb$ satisfying  $\Upsilon'' \geq \Upsilon$ and $\Upsilon'' \geq \Upsilon'$. Thus  $(\Lb,\geq)$ is a directed set.

\subsection{Combined Holonomy and Background exponential structures}\label{sec2C}

The combined set of labels associated to holonomies and background exponentials is given by pairs $l=(\gamma,\Upsilon) \in \Lh \times \Lb =: \L$ with preorder relation given by $(\gamma',\Upsilon') \geq (\gamma,\Upsilon)$ iff $\gamma' \geq \gamma$ and $\Upsilon' \geq \Upsilon$. $\L$ is then a directed set, which will be used in section \ref{sec5} to construct the projective limit description of the KS quantum configuration space.  

Given $l=(e_1,\ldots,e_n,E_1,\ldots,E_N) \in \L$ we define the group
\be
G_l := SU(2)^n \times U(1)^N \label{gl}
\ee
and the map
\ba
\pi_l: \A & \to &  G_l \\
A & \mapsto & \pi_l[A] := (h_{e_1}[A],\ldots,h_{e_n}[A], \beta_{E_1}[A],\ldots ,\beta_{E_N}[A]). \label{pil} 
\ea
Most of the results in the present work rely on the result that $\pi_l[\A] \subset G_l$ is dense in $G_l$ for any label $l \in \L$.  This is shown in section \ref{sec3} and appendix \ref{applemma}.  

\subsection{KS representation} \label{sec2D}

The KS Hilbert space, $\Hks$,  is spanned by states of the form $|s,{\Eb}\rangle$, where $s$ is an LQG spin network and $\Eb$ a background electric field. The  inner product is given by
\begin{equation}
\langle s^{\prime},{\Eb}^{\prime }_{}|s,{\Eb}\rangle = \bra s|s^{\prime} \ket_\lqg
\delta_{{\Eb}^{\prime}, {\Eb}}, \label{ksip}
\end{equation}
where $\bra s|s^{\prime} \ket_\lqg$ is the spin network LQG inner product and $\delta_{{\Eb}^{\prime}, {\Eb}}$ the Kronecker delta.

Holonomies (\ref{hol}) and background exponentials (\ref{baexp}) act by
\begin{eqnarray}
{\hat h}^{\phantom{e\;}C}_{e\;D}|s,\Eb \rangle &=& \vert {\hat h}^{\lqg \,A}_{e\; \quad B} s,\Eb_{}\rangle ,
\label{holhat}
\\
{\hat \beta}_{\Eb'} |s,\Eb \ket &=&  |s, \Eb'+\Eb \ket \label{betahat}.
\end{eqnarray}
Above, we have used the notation of \cite{me} wherein 
given an LQG operator ${\hat O}$ with action 
$\hat O|s\rangle = \sum_IO^{(s)}_I|s_I\rangle$  in standard LQG, we have defined the state $|{\hat O}s, { E}\rangle$
in the KS representation through
\begin{equation} 
|{\hat O}s ,{E}_{}\rangle:= \sum_IO^{(s)}_I|s_I,{ E}_{} \rangle .
\label{oks}
\end{equation}
The action of fluxes is given by 
\be
{\hat F}_{S,f}|s,\Eb \rangle = |{\hat F}^\lqg_{S,f} s,\Eb \rangle + F_{S,f}(\Eb)|s,\Eb \rangle ,
\label{fluxhat}
\ee
where $f^i$ is the $su(2)$-valued smearing scalar on the surface $S$ and $F_{S,f}(E) = \int_S dS_a f^i E^a_i$ the  flux associated to the background electric field $E^a_i$.

The KS representation supports a unitary action of spatial diffeomorphisms and gauge transformations which can be used to construct a diffeomorphism and $SU(2)$ gauge invariant space via group averaging techniques  \cite{hs,miguelme}.

\section{The Master Lemma} \label{sec3}

In this  section we state the Master Lemma and describe the idea behind its proof. We conclude with a summary of the 
steps in the proof. These steps are implemented in detail in  appendix \ref{applemma}.\\

\noindent{\bf Statement of the Lemma}:\\
\noindent Let 
$e_1,\ldots,e_n$ be $n$ independent edges. Let 
$E_1,\ldots E_N$ be $N$ rationally independent semianalytic $SU(2)$ electric fields.
Let ${\rm G}$ be the product group ${\rm G}:= SU(2)^n \times U(1)^N$.
Define the map $\phi: \A  \to  {\rm G}$  through
\be
\phi(A) := (h_{e_1}[A],\ldots,h_{e_n}[A], \beta_{E_1}[A],\ldots ,\beta_{E_N}[A]). \label{phiA}
\ee
Then the image $\phi (\A )$ of $\A$ is dense in ${\rm G}$.
\\
\noindent{\bf Idea behind the Proof}:\\
Let ${\rm g}\in {\rm G}$ so that   
${\rm  g}= (g_1, \ldots, g_n, u_1, \ldots, u_N)$ where $g_{\alpha} \in SU(2),\; \alpha =1,..,n$ and
$u_I \in U(1),\; I=1,..,N$. Then it suffices to show that for any given ${\rm g}\in {\rm G}$
and any $\delta > 0$, there exists an element $A^{{\rm g},\delta}\in \A$ such that 
\ba
|h_{e_{\alpha}}[A^{{\rm g},\delta}]_{\;C}^{\;\;D}- g_{\alpha C}^{\;\;D}| &\leq & C_1 \delta\; \forall \;\;\alpha=1,..,n\;  {\rm and}\; 
C,D=1,2.
\label{hgdelta}
\\
|\beta_{E_I}[A^{{\rm g},\delta}] - e^{i \theta_I}| & \leq & C_2 \delta \;\forall\; \;I=1,..,N,
\label{budelta}
\ea
where $C_1,C_2$ are $\delta$- independent constants, $C,D$ are $SU(2)$ matrix indices and 
$u_I =: e^{i \theta_I}\;, \theta_I\in \reals$. 
We shall show that the above equations hold with $C_1= 0$ and an appropriate choice of $C_2$.
In what follows we shall drop the superscript ${\rm g}$ in $A^{{\rm g},\delta}$ to avoid notational clutter.

First we construct a connection $A^{B,\delta}$ which satisfies equation (\ref{budelta}). This is done using the 
rational independence of the set of electric fields in conjunction with standard results on the Bohr compactification
of $\reals^m$ \cite{hewittross}). In general, of course, the evaluation of the edge holonomies on this connection will not satisfy
equation (\ref{hgdelta}).

On the other hand, from standard LQG results \cite{aajurekhoop}, given {\em any} set of $n$ group elements
we are guaranteed the existence of
a connection whose holonomies along the $n$ independent edges $\{e_{\alpha} \}$ reproduce these group elements
{\em exactly}. Further, these LQG results imply that such a connection $A^{\epsilon}$  can be constructed for any positive
$\epsilon$ such that it vanishes
everywhere except around balls of radius $\epsilon$, each such ball intersecting the interior of each edge in 
an $\epsilon$ size segment (where $\epsilon$ is a coordinate distance, measured in fixed coordinate charts). 
Moreover, since the connection samples only an $\epsilon$ size segment of each edge, it can be shown that the 
connection is of order $1/\epsilon$. Clearly the {\em three dimensional integral} of  such a  connection 
yields order $\epsilon^2$ contributions. 

Were we to add such a connection to $A^{B,\delta}$ above, then, for small enough $\epsilon$, it would have a negligible
effect on the conditions (\ref{budelta}). However, the connection $A^{B,\delta}$ has, in general, support on the 
set of edges $\{e_{\alpha}\}$ and hence contributes to the edge holonomies. The idea then is to carefully choose the connection 
$A^{\epsilon}$ so that its contributions together with those from $A^{B,\delta}$ yield the set of elements 
$\{g_{\alpha}\}$. In order to do this we need to cleanly seperate the contributions of $A^{B,\delta}$ 
from those of $A^{\epsilon}$.
This would be easy to do if 
$A^{B,\delta}$ and $A^{\epsilon}$ had mutually exclusive supports; if this were so 
the integral over each edge $e_{\alpha}$  of  $A^{B,\delta}+ A^{\epsilon}$ would seperate into
contributions over
segments of this edge where each segment supports either $A^{B,\delta}$ or $A^{\epsilon}$ but not both.
We could then write each 
edge holonomy of $A^{B,\delta}+ A^{\epsilon}$ in terms of  
compositions of holonomies along the segments of each edge, each  segment holonomy being
evaluated solely with respect to $A^{B,\delta}$ or solely 
with respect to $A^{\epsilon}$. We could then choose $A^{\epsilon}$ so as to ``undo'' the contributions from
$A^{B,\delta}$ and yield the required group elements $g_{\alpha}$

Indeed, as shown in the Appendix, we can choose the supports of $A^{B,\delta}$
and $A^{\epsilon}$ such that each edge $e_{\alpha}$ can be written as $s^1_{\alpha}\circ s_{\alpha} \circ s^2_{\alpha}$
with  $\s_{\alpha}$ in the support of $A^{\epsilon}$ and $\s^1_{\alpha}, \s^2_{\alpha}$ in the support of 
$A^{B,\delta}$ so that $h_{e_{\alpha}}[A^{B,\delta}+ A^{\epsilon}]$ takes the form 
$h_{s^1_{\alpha}}[A^{B,\delta}]h_{s_{\alpha}}[A^{\epsilon}]h_{s^2_{\alpha}}[A^{B,\delta}]$.
We then choose  $A^{\epsilon}$ such that 
$h_{s_{\alpha}}[A^{\epsilon}]= (h_{s^1_{\alpha}})^{-1}g_{\alpha}(h_{s^2_{\alpha}})^{-1}$ so that conditions 
(\ref{hgdelta}) are satisfied with $C_1=0$.

To obtain $A^{B,\delta}$ with the desired support we first construct a  connection ${\bar A}^{B, \delta}$ which satisfies
the conditions (\ref{budelta}) and then multiply it with a semianalytic function of appropriate support.
To do so, recall that the support of $A^{\epsilon}$ is in balls of size $\epsilon$. We construct
a ball of size $2\epsilon$ around each such ball. Then the desired function is constructed so as to
equal unity outside these 
balls of size $2\epsilon$, and vanish inside the $\epsilon$ size balls which support $A^{\epsilon}$.
Since the modification of ${\bar A}^{B, \delta}$ is only in regions of 
 order $\epsilon^3$, for small enough $\epsilon$  these modifications contribute negligibly to the background exponentials
and one can as well use $A^{B,\delta}$ instead of ${\bar A}^{B, \delta}$ to satisfy the conditions (\ref{budelta}).

The technical implementation of the proof then proceeds along the following steps which are detailed in Appendix \ref{applemma}:\\

\noindent (i) Using standard results from Bohr compactification of $\reals^m$, 
we construct a connection ${\bar A}^{B, \delta}$ which satisfies (\ref{budelta}) for some $C_2$.\\

\noindent (ii)  For sufficiently small $\epsilon$ and for appropriately chosen $\epsilon$- independent charts,
we show the existence of balls $B_{\alpha}(2\epsilon ), \alpha =1,..,n$ of coordinate size  $2\epsilon$ such that 
\ba
B_{\alpha}(2\epsilon ) \cap B_{\beta}(2\epsilon ) = \emptyset \;\; {\rm iff} \;\; \alpha \neq \beta && \\
B_{\alpha}(2\epsilon ) \cap \et_{\beta} = \emptyset \;\; {\rm iff} \;\; \alpha \neq \beta &&\\
\bar{B}_{\alpha}(2\epsilon ) \cap \et_{\alpha} \;\; {\rm is}\;\; {\rm a}\;\; {\rm  semianalytic}\;\; {\rm  edge}&&
\label{2epsilonedge}
\ea

\noindent (iii) We construct a real semianalytic function $f_{\epsilon}$ such that $|f_{\epsilon}|\leq 1$ on $\Sigma$
with 
\ba 
f_{\epsilon}&=& 1\;\; {\rm on}\;  \Sigma - \cup_{\alpha} B_{\alpha}( 2\epsilon ) \\
            &=& 0 \;\; {\rm on}\;  \cup_{\alpha} B_{\alpha}( \epsilon ),\\
\ea
where $B_{\alpha}(\epsilon )$  denotes the $\epsilon$ size ball with the same centre as $B_{\alpha}(2\epsilon )$.\\

\noindent (iv) 
From (\ref{2epsilonedge}) it follows that 
\be
e_{\alpha} = s^1_{\alpha} \circ s_{\alpha} \circ s^2_{\alpha}
\ee
with 
\ba
\tilde{s}_{\alpha} &:= & \et_{\alpha} \cap {\bar B}_{\alpha}(\epsilon )\\
 \tilde{s}^1_{\alpha} \cup \tilde{s}^2_{\alpha} & = & \et_{\alpha} \cap (\Sigma -{B}_{\alpha}(\epsilon )) .
\ea
Define:
\be
h_{s^i_{\alpha}}[A^{B,f}] =: g^i_{\alpha}, \;\;i=1,2
\ee
where $A^{B,f}:= f_{\epsilon} {\bar A}^{B,\delta}$.
Then we construct a connection $A^{\epsilon}$ supported in $\cup_{\alpha} B_{\alpha}(\epsilon )$ such that 
\be
h_{s_{\alpha}}[A^{\epsilon}] = (g^1_{\alpha})^{-1}g_{\alpha}(g^2_{\alpha})^{-1} \label{step4}
\ee

\noindent (v) We define $A^{\delta}:= A^{B,f} + A^{\epsilon}$ and show that conditions 
(\ref{hgdelta}) and (\ref{budelta}) are satisfied with $C_1=0$ and some $C_2$.

\section{$C^*$- Algebraic Considerations} \label{sec4}
\subsection{$C^*$ algebra $\hbabar$}
We denote by $\hba$ the $*$-algebra of functions of $\A$ generated by the elementary functions (\ref{hol}) and (\ref{baexp}), with * relation given by complex conjugation. A generic element of $\hba$ takes the form:
\be 
a[A]= \sum_{i=1}^M  c_i  \beta_{E'_i}[A] h_{e'^i_1}[A]_{C^i_1}^{\; D^i_1} \ldots h_{e'^i_{m_i}}[A]_{C^i_{m_i}}^{\; D^i_{m_i}}  \in \hba, \label{psi} 
\ee
for given $M$ complex numbers $c_i$, $M$ electric fields $E'_i$,  $\sum_{i=1}^{M} m_i$ edges $e'^i_k$, and choices of matrix elements for the $SU(2)$ holonomies, $C^i_k, D^i_k \in \{1,2\}$. Since holonomies and background exponentials are bounded functions of $\A$, elements of $\hba$, are bounded. Thus, the sup norm is well defined on  $\hba$:
\be
|| a || := \sup_{A \in \A} |a[A]|,  \quad a \in \hba .\label{algnorm}
\ee
Being a sup norm, it is compatible with the product and complex-conjugation star relations, so that upon completion we obtain a  unital $C^*$ algebra denoted by $\hbabar$.  By Gel'fand theory $\hbabar$ can be identified with the $C^*$ algebra of continuous functions on a compact, Hausdorff space $\spec$,  $\hbabar \simeq C(\spec)$. It will be useful for later purposes to denote by $\cyl(\spec) \subset C(\spec)$ the subalgebra of continuous functions corresponding to $\hba$ in the Gel'fand identification. $\cyl(\spec)$ will be referred to as the space of cylindrical functions of $\spec$. 

Finally, the fact that $\hba$ separates points in $\A$ implies that $\A$ is topologically  dense in $\spec$ \cite{rendall}. Hence $\spec$ represents a  space of generalized connections.

\subsection{Characterization of elements of $\hba$} \label{charhba}

It will  be useful to characterize elements of $\hba$ by identifying independent edges and electric fields involved in any given algebra element as follows.

Let $a \in \hba$ so that it is of the form (\ref{psi}).

Let $(E_1, \ldots, E_N) \;,  N \leq M,$ be a set of independent electric fields as defined in section \ref{sec2}
in terms of which all  $E'_J$'s in (\ref{psi}) can be written as integral linear combinations:
\be
E'_J = \sum_{I=1}^{N} k^{I}_{J} E_{I}, \quad I=1,\ldots,N, \quad k^I_J \in \integers \label{genset}
\ee
(such an algebraically independent generating set always exists, see Appendix \ref{gensetapp}). 
From (\ref{genset}) it follows that  
the background exponentials in (\ref{psi}) can be replaced by appropriate products of background exponentials 
(and their complex conjugates) associated with these  independent electric fields. 

Let $(e_1,\ldots,e_n), \; n \leq \sum_{i=1}^{M} m_i$ be a set of independent edges as defined in section \ref{sec2}  such that all edges in (\ref{psi}) can be obtained as compositions of them or their inverses.
It follows that 
the edge holonomies in (\ref{psi}) can be replaced by products of holonomies (and their complex conjugates) along 
these independent edges.

With these replacements the element $a$ acquires the form of a polynomial in the  holonomies along the independent edges,
background exponentials associated with the independent electric fields, and their complex conjugates.
Thus we have shown that with 
$l:=(e_1,\ldots,e_n,E_1, \ldots, E_N), l\in \L$,  the algebra element (\ref{psi}) takes the form: 
\be
a[A] = a_l(\pi_l[A]), \label{Ff}
\ee
where $\pi_l[A] \equiv (h_{e_1}[A],\ldots,h_{e_n}[A], \beta_{E_1}[A],\ldots ,\beta_{E_N}[A])\in G_l \equiv  SU(2)^n \times U(1)^N $ as in Eq. (\ref{pil}) and  $a_l: G_l \to \complex$ is
a function that depends polynomially on 
the $SU(2)$ and $U(1)$ entries and their complex conjugates.   
Clearly, there exist many choices of $l$ for which equation (\ref{Ff}) holds. 

It is then useful to define the notion of {\em compatibility} of $a$ and $l$. 
Given $a \in \hba$ so that $a$ is necessarily of the form (\ref{psi}), let $l$ be such that all edges  
and all electric fields  in (\ref{psi}) can be obtained in terms of compositions of edges and integral linear
combinations of electric fields in $l$. Then  we shall say that $l$
is {\em compatible} with $a$, that $a$ is {\em compatible} with $l$ and that $a,l$ are {\em mutually compatible}.
In this language what we have shown above is that given $a \in \hba$ and $l$ which is compatible with $a$,
there exists a function $a_l: G_l \to \complex$ with polynomial dependence on its  $SU(2)$ and $U(1)$ entries
and their complex conjugates such that (\ref{Ff}) holds.

A key result which we shall need, and which  follows directly from the lemma may then be stated as follows:\\
Let $l \in \L$ and $a\in \hba$ such that $l$ is compatible with $a$. Then the function  $a_l$ as constructed above is the
{\em unique} continuous function on $G_l$ whose restriction to 
$\pi_l[\A] \subset G_l$ agrees with $a[A]$.\footnote{The result follows immediately from the fact that
$a_l$ is manifestly continuous on $G_l$, that Eq. (\ref{Ff}) gives the values of $a_l$ on the set 
$\pi_l[\A] \subset G_l$ and that, by the lemma of section \ref{sec3}, this is a dense subset of $G_l$.}
Accordingly, we shall say that (given mutually compatible $a,l$)
$a_l$ is {\em uniquely determined} by $a,l$.


Next, note that if $l$ is compatible with $a$ and $l''$ is such that $l''\geq l$, then $l''$ is compatible with 
$a$ so that we have  
 \be
a[A]=  a_{l''}(\pi_{l''}[A]) = a_l( \pi_l[A])\label{apil}.
 \ee
It is of interest to elucidate the relationship between 
$a_l$ and $a_{l''}$. We proceed as follows.

 Since $l''\geq l$,  edges $e_{i} \in l$ can  be written as compositions of edges in $l''$. Let us denote this relation by: $e_i= \tilde{p}_i(e''_1,\ldots )$, where $\tilde{p}_i$ denotes a particular composition of edges (and their inverses) in $l''$. This corresponds to a relation on the holonomies of the form:
\be
h_{e_i}[A] = h_{\p_i(e''_1,\ldots)}[A] = p_i(h_{e''_1}[A], \ldots ), \label{hepp}
\ee
where  $p_i: SU(2)^{n''} \to SU(2)$  is the map determined by interpreting the compositions rules of $\tilde{p}_i$ as matrix multiplications. For example, if $e_1=e''_2 \circ  (e''_1)^{-1}$ then $p_1(g''_1,\ldots,g''_{n''})=g''_2 (g''_1)^{-1}$.
Similarly,  electric fields $E_I \in l$ can be written as integer linear combinations of electric fields in  $l''$:
\be
E_I = \tilde{P}_I (E''_1 ,\ldots) :=  \sum_{J=1}^{N''} q_{I}^{J} E''_{J}, \quad q_{I}^{J} \in \integers, \quad I=1,\ldots,N. \label{Ptilde}
\ee
Associated to (\ref{Ptilde}) there is the map $P_I: U(1)^{N''} \to U(1)$ given by $P_I(u''_1,\ldots,u''_{N''})= \Pi_{J=1}^{N''}(u''_J)^{q_{I}^{J}}$ so that
\be
\beta_{E_I}[A] = \beta_{ \tilde{P}_I (E''_1 ,\ldots)}[A] = P_I(\beta_{E''_1}[A], \ldots ). \label{bepp}
\ee 
The above maps combine in a map\footnote{This map will be of later use in the projective limit construction of section \ref{sec5}.} 
\be
p_{l , l''}:=(p_1,\ldots,p_n,P_1,\ldots,P_N): G_{l''} \to G_l \label{projmap}
\ee
(that is `block diagonal'  in the $SU(2)$ and $U(1)$ entries). Equations (\ref{hepp}) and (\ref{bepp}) can then be summarized as:
\be
\pi_l[A] =p_{l , l''}( \pi_{l''}[A]).\label{pip}
\ee
Substituting (\ref{pip}) in the last term of (\ref{apil}) we find:
\be
a_{l''}(\pi_{l''}[A]) = a_l(p_{l,l''}(\pi_{l''}[A])). \label{alpp}
\ee
Thus $a_{l''}$ and $a_l \circ p_{l , l''}$ coincide on the dense subset $\pi_{l''}[\A] \subset G_{l''}$. Since $a_{l''}$ and $a_l \circ p_{l , l''}$ are continuous functions on $G_{l''}$ we conclude that
 \be
 a_{l''} = a_{l} \circ p_{l , l''} \label{rellabels}.
 \ee

We conclude this section by noting one more consequence of the Lemma:\\
\noindent The algebra norm (\ref{algnorm}) of $a \in \hba$  as in (\ref{Ff}) coincides with the sup norm on $G_l$ of $a_l$:
\be 
|| a ||= \sup_{A \in \A} |a_l(\pi_l[A]) | = \sup_{g \in G_l} |a_l(g)|.\label{normF}
\ee

\subsection{Characterization of $\spec$} \label{sec4C}
 
One of the characterizations of the quantum configuration space in standard LQG is given by the set $\hom(\P,SU(2))$ of homomorphisms from the path groupoid $\P$ to $SU(2)$  \cite{velhinho,ttbook}.  The analogue space associated to the background exponentials is given by $\hom(\E,U(1))$, the set of homomorphisms from the abelian group $\E$ of semianalytic electric fields (with abelian product  given by addition) to $U(1)$.

We will now  establish a one-to-one correspondence between $\spec$ and $\hom(\P,SU(2)) \times \hom(\E,U(1))$.

First we show that any element $\phi \in \spec$ defines an element of $\hom(\P,SU(2)) \times \hom(\E,U(1))$.
Recall that from Gel'fand theory, $\phi$ is a $C^*$ algebraic homomorphism from the $C^*$ algebra $\hbabar$
to the $C^*$ algebra of complex numbers $\complex$. Let $\ha$ and $\Ba$ be the $*$- algebras generated, 
respectively, by only the holonomies and by only the background exponentials. The sup norm on $\hba$ defines
a  norm on each of  $\ha$ and $\Ba$ and the  two algebras can then be completed in their norms so defined
to yield the $C^*$ algebras $\habar$ and $\Babar$. Clearly $\habar$ and  $\Babar$ are subalgebras of 
$\hbabar$ with $\habar$ being exactly the holonomy $C^*$ algebra of LQG.


Let $\phi_{\rm H}:=\phi|_{\habar}$ be the restriction of $\phi$ to $\habar$. $\phi_{\rm H}$ is a homomorphism from $\habar$ to $\complex$. 
By the standard LQG description \cite{velhinho,ttbook}, $\phi_{\rm H}$ defines an element $s_\phi \in \hom(\P,SU(2))$ given by
\be
s_\phi(p)_{C}^{\;D} := \phi(h_{p\; C}^{\; \;\;D}).
\ee

Similarly, it is easy to verify that  $\phi_{\text{B}}:=\phi|_{\Babar}$ defines an element $u_\phi \in \hom(\E,U(1))$ given by 
\be
u_\phi(E):= \phi(\beta_E),
\ee
since $u_\phi(0)=1$, $u_\phi(E_1+E_2)=u_\phi(E_1)u_\phi(E_2)$ and $\overline{u_\phi(E)}=u_\phi(-E)$, implying $u_\phi(E) \in U(1)$ (see \cite{vel2,ttbook} for the analogue statement in the context of $\reals$-Bohr.)

Conversely, given $s \in \hom(\P,SU(2))$ and $u \in \hom(\E,U(1))$ we want to find  $\phi \in \spec$ such that $u_\phi=u$ and $s_\phi=s$. Following the same strategy as in LQG \cite{velhinho,vel2,ttbook},  we first find a  homomorphism $\phi : \hba \to \complex$, and then show it is bounded and hence extends to $\hbabar$. 

Given a general element $a \in \hba$, it can always be written in the form (\ref{Ff}) for 
any $l$ compatible with $a$. Accordingly, we choose some compatible 
$l=(e_1,\ldots,e_n,E_1, \ldots, E_N)$ and  define $\phi$ on $\hba$ by:
\be
\phi(a) := a_l(s(e_1),\ldots,s(e_n),u(E_1),\ldots,u(E_N)) \label{defhom}.
\ee
Since $a_l$ is uniquely defined (see section 4B), there is no ambiguity in this definition
if $l$ is specified. However, there are infinitely many $l$ which are compatible with $a$. We now show
that $\phi (a) \in \complex$ given by (\ref{defhom}) is {\em independent} of the choice of such $l$.
Accordingly, let $l',l$ be compatible with $a$. Then we need to show that 
\be
a_{l}(s(e_1),\ldots,u(E_1),\ldots)= a_{l'}(s(e'_1),\ldots,u(E'_1),\ldots). \label{homwelldef}
\ee
This can be shown by writing $a_{l}$ and $a_{l'}$ in terms of a finer label $l''$ such that $l'' \geq l'$ and $l'' \geq l$ according to (\ref{rellabels}), and using the homomorphism properties of $s$ and $u$. Let $p_i : SU(2)^{n''} \to SU(2)$ and $P_I :=U(1)^{N''} \to U(1)$ be the maps described in section \ref{charhba} determined by the way probes of $l$ are written in terms of those of $l''$. The homomorphism property of $s$ and $u$ imply that 
\ba
s(e_i) & = & p_i(e''_1,\ldots) \\
u(E_I) & =& P_I(E''_1,\ldots),
\ea
substituting these relations in the RHS of (\ref{defhom}) and using the result (\ref{rellabels}) we find
\be
a_l(s(e_1),\ldots,u(E_1),\ldots) = a_{l''}(s(e''_1),\ldots,u(E''_1),\ldots) .
\ee
Repeating the argument for the set $l'$, one concludes
\be
a_{l'}(s(e'_1),\ldots,u(E'_1),\ldots) = a_{l''}(s(e''_1),\ldots,u(E''_1),\ldots) .
\ee
Hence (\ref{homwelldef}) follows and (\ref{defhom}) is  independent of the choice of compatible $l$.  

Next we show that $\phi$ so defined is a homomorphism to $\complex$.
By choosing any fixed $l$ it trivially follows that:\\
\noindent (a) $\phi$ maps the zero element of $\hba$ to $0\in \complex$.\\
\noindent (b) $\phi$ maps the unital element of $\hba$ to $1\in \complex$. \\
\noindent (c) given any complex number $C$ and algebra element $a\in \hba$, $\phi (Ca)= C\phi (a)$.\\
Further note that there exists a `fine enough' $l$ which is simultaneously
compatible with a given set of elements  $a, b, ab, a+b, a^* \in \hba$.
From the continuity of $a_l, b_l, (ab)_l, (a+b)_l, (a^*)_l$ on $G_l$, the continuity preserving
property of the 
operations of addition, multiplication and complex conjugation on the space of continuous functions
on $G_l$ and the uniqueness of the specification of any $c_l:G_l\rightarrow \complex$ given mutually compatible
$c\in \hba, l\in \L$ (see section 4B), it follows that:\\
\noindent (d) $\phi (ab) = \phi (a) \phi (b)$, $\phi (a+b) = \phi (a) + \phi (b)$ and $\phi(a^*)=\phi(a)^*$. \\
Properties (a)-(d) show that  equation (\ref{defhom}) defines a homomorphism from $\hba$ to $\complex$.
Next, we note that due to the Lemma and Eq. (\ref{normF})  $\phi$ is bounded  since:
\be
|\phi(a)| = |a_l(s(e_1),\ldots,u(E_1),\ldots)|  \leq  \sup_{g \in G_l} |a_l(g)| = ||a||. 
\ee 
It then follows that $\phi$ uniquely extends to a homomorphism from $\hbabar$ to $\complex$ (see  \cite{ttbook} around  Eq. 6.2.71 for discussion of extension of bounded homomorphisms to completed algebra).

Finally it is easy to verify explicitly that (\ref{defhom}) 
satisfies $u_\phi=u$ and $s_\phi=s$, thus establishing the correspondence between 
$\spec$ and $\hom(\P,SU(2)) \times \hom(\E,U(1))$.

 \subsection{Realization of the KS Hilbert space as the space $L^2(\spec,\muks)$} \label{sec4D}
 

We now use the  KS representation of $\hba$ (see section \ref{sec2D} and  \cite{miguelme}) to construct a positive linear functional (PLF) on $\hbabar$. Given $a \in \hba$, let us denote by $\hat{a}$ the corresponding operator in the KS Hilbert space $\Hks$. We define the PLF  by
\be
\w(a):= \bra 0,0 | \hat{a} | 0,0  \ket , \label{defplf}
\ee
where $|0,0 \ket \in \Hks$ is the  KS 'vacuum' state corresponding to the trivial spin network and vanishing background electric field. As in  LQG \cite{ttbook,velhinho,vel2},  the PLF can be written as an integral over the group elements: For $a \in \hba$ given by (\ref{Ff}), we have (see  appendix \ref{plfintapp} for a proof):
\be
\w(a)= \int_{G_l}  a_l(g) d \mu_l ,\label{plfint}
\ee
for any  $l\in \L$ compatible with $a$. Here $d \mu_l$ is the Haar measure on the group $G_l$ normalized so that $\int_{G_l} d \mu_l =1$.
Boundedness of $\w$ follows from the Lemma via Eq. (\ref{Ff}):
\be
| \w(a)| =\left| \int_{G_l} a_{l}(g) d \mu_l \right| 
 \leq   \sup_{g \in G_l}  |a_l(g)| = ||a|| .
 \ee
Thus $\w$ uniquely extends to  $\hbabar \simeq C(\spec)$ \cite{ttbook}. The Riesz-Markov theorem then implies the existence of a regular measure $\muks$ on $\spec$ such that:
\be
\w(a)=\int_{\spec} a d \muks, \label{riesz}
\ee
where in the RHS of (\ref{riesz}) $a$ is seen as a an element of  $C(\spec)$ via the Gel'fand identification. By construction it follows that $\Hks \simeq L^2(\spec,\muks)$, since $\Hks$ can be identified with the GNS Hilbert space associated with $\w$, and the two constructions lead to the same representation \cite{velhinho}.  

Elements $a \in \hbabar \simeq C(\spec)$ have now a dual interpretation: When seen as elements of $C(\spec)$ we will interpret them as `wavefunctions' in the $L^2$ representation, i.e. \emph{vectors} on the Hilbert space. When seen as elements of $\hbabar$, we will usually associate them to  \emph{operators} $\hat{a}$ on the Hilbert space $\Hks$. In the `wavefunction' picture, $\cyl(\spec)\simeq \hba$ plays a special role: It is a dense subspace of $L^2(\spec,d\muks)$ which serves as a dense domain for the definition of the  unbounded flux operators ${\hat F}_{S,f}$. In the next section we discuss the action of fluxes (\ref{fluxhat}) in this `wavefunction' picture. 
\subsection{Action of Fluxes on $L^2(\spec,\muks)$} \label{sec4E}

In the $L^2$ description of  $\Hks$, the  KS spinnet $|s,E \ket$ corresponds to the `wavefunction'  
$T_s \beta_E \in \cyl(\spec)$, where $T_s[A] \in \hba$ is the spin network function associated to  $s$ \cite{ttbook}, 
$\beta_E[A] \in  \hba$ the background exponential function (\ref{baexp}), and $T_s,\beta_E$ the respective elements in 
$\cyl(\spec)$ under the Gel'fand identification $\hba \simeq \cyl(\spec)$. 

Since the action of the flux operator ${\hat F}_{S,f}$ on $|s,\Eb \rangle$   yields the finite linear combination of KS spinnets (\ref{fluxhat}), we can translate this action as a map ${\hat F}_{S,f} : \cyl(\spec) \to \cyl(\spec)$. In this description, Eq. (\ref{fluxhat}) takes the form
\be
{\hat F}_{S,f}( T_s \beta_E)= 
({\hat F}^{\lqg}_{S,f}T_s) \beta_E+ F_{S,f}(E) T_s \beta_E.  \label{fluxwf}
\ee
Here $({\hat F}^{\lqg}_{S,f}T_s) \in \cyl(\spec)$ 
denotes the finite linear combination of spin networks, ${\hat F}^\lqg_{S,f}| s \ket$,  obtained by the action of 
the flux operator labelled by $(S,f)$ in the standard LQG representation. 
Recall that since $T_s[A] \in \hba$, it is a polynomial in a set of independent edge holonomies.
Using the 
algebraic identification of $\hba$ with $\cyl(\spec)$, it then follows from standard LQG 
that the correspondent in $\hba$ of $({\hat F}^{\lqg}_{S,f}T_s) \in \cyl(\spec)$ is 
$- i \Xh_{S,f}(T_s[A])\in \ha \subset \hba$ where $\Xh_{S,f}$ is a `derivative operator' whose action on $T_s[A]$
is built out of that of 
left and right invariant
vector fields of $SU(2)$, 
on the $SU(2)$ valued edge holonomies  underlying $T_s[A]$ \cite{area}. 

It is useful for the purposes of section \ref{sec6}, to note that we may also re-express 
$\Xh_{S,f}(T_s[A])$ as the classical Poisson bracket $\{ T_s[A],F_{S,f}\}$ 
 (see for example References
\cite{acz,ttbook} as well as equation (\ref{xs}) below). This sort of re-expression extends to
both the terms in the right hand side of equation (\ref{fluxwf}) so that the correspondent in 
$\hba$ of the right hand side can be written as the Poisson bracket $-i \{ T_s[A] \beta_E[A], F_{f,S} \}$.

Next note that since {\em any} element of $\hba \simeq \cyl (\spec )$ is a polynomial in the holonomies and
background exponentials, the Peter-Weyl theorem (see for example \cite{ttbook})
implies that any such element can be re-expressed  as a finite linear combination of KS spin net functions
i.e. $a[A]$ can be written as an expansion $\sum_i c_i T_{s_i}[A]\beta_{E_i}[A]$ for suitably defined spinnets $s_i$
and electric fields $E_i$. Denoting the wave function in $\cyl (\spec )$ corresponding to $a[A]\in \hba$ by 
$a$, it then follows from the previous paragraph and equation (\ref{fluxwf}) 
that the $\hba$ correspondent of ${\hat F}_{S,f}a$ is $-i \{ a[A], F_{S,f} \}$.\footnote{This result can also be interpreted as follows. Since the KS representation is a representation of the classical Poisson algebra \cite{miguelme}, it follows that given $a[A] \in \hba$ and $\{a[A] ,F_{S,f} \} \in \hba$, we have that $[ \hat{a},{\hat F}_{S,f} ]=i \widehat{\{a ,F_{S,f} \}} $. The `wavefunction' associated to 
$a \in \hba$ corresponds to the vector $\hat{a} | 0 \ket \in \Hks$ where $|0 \ket$ is the KS vacuum. Using the fact 
that ${\hat F}_{S,f} |0 \ket =0$ and the above commutation relation, one concludes that the $\hba$ element 
associated to the wave function $({\hat F}_{S,f}a)$ is given by $-i \{a[A] ,F_{S,f} \}$.}


\section{The quantum configuration space as a projective limit}\label{sec5}

As in LQG, the quantum configuration space $\spec$ admits a characterization as a projective limit space. In section \ref{sec5A} we  describe the projective limit space,  denoted by $\Abar$,  and show that it is homeomorphic to $\spec$.  In section \ref{sec5B} we discuss measure theoretic aspects of $\Abar$.

\subsection{Topological identification of $\Abar$ with $\spec$} \label{sec5A}
The ingredients in the construction of $\Abar$ are: i) The directed set $\L$ and the family of compact spaces $\{ G_l , l \in \L\}$ defined in section \ref{sec2C};  ii) The continuous projections $p_{l,l''}: G_{l''} \to G_{l}, \; l'' \geq l$ described  in section \ref{charhba},  Eq. (\ref{projmap}). 

Recall that $p_{l,l''}$ is determined by the way probes in $l$ are written in terms of probes in $l''$. These maps are surjective\footnote{The master lemma implies that $p_{l,l'}(G_{l'})$ is dense in $G_l$. Continuity of $p_{l l'}$ implies compactness of  $p_{l,l'}(G_{l'})$. Since $G_l$ is Hausdorff $p_{l,l'}(G_{l'})$ is closed and hence $p_{l,l'}(G_{l'})=G_l$.} and it is easy to verify that  if $l'' \geq l' \geq l$ then $p_{l,l''}=p_{l,l'} \circ p_{l',l''}$. Thus $( \L,\{ G_l \} ,\{ p_{l l'} \})$  satisfy the required conditions for the construction of a  projective limit space   \cite{aajurekproj,encytop}.


Let us describe the main features of $\Abar$ (see Appendix \ref{projlimapp} for additional details as well as for a comparison with the usual construction in LQG). A point in $\Abar$ is given by an  assignment of points  $x_l \in G_l$ for each $l \in \L$ satisfying the consistency condition $x_l = p_{l,l'}(x_{l'})$ whenever $l' \geq l$.  
$\Abar$ is a compact Hausdorff space (see \cite{ttbook} and Appendix \ref{projlimapp}). 
We denote by $\{ x_{l'} \}$ an element of the projective limit space. The  canonical projections
\ba
p_l : & \Abar & \to  G_l  \label{pl} \\
& \{ x_{l'} \} & \mapsto  x_l 
\ea
satisfy $p_l= p_{l,l'} \circ p_{l'}$ for $l' \geq l$, are continuous,\footnote{The topology of $\Abar$ corresponds to the weakest topology such that the maps (\ref{pl}) are continuous \cite{ttbook}.} and as shown in \cite{aajurekproj} surjective.\footnote{We thank Jos\'e Velhinho for pointing us to reference \cite{aajurekproj}.}


Note that given a  connection $A \in \A$, 
the points $\pi_l[A] \in G_l$  satisfy the projective consistency conditions 
 with respect to the 
projections $p_{l,l'}$  by virtue of Eq. (\ref{pip}). 
This implies that every $A \in \A$ defines an element of 
$\Abar$. Since holonomies and background exponentials seperate points in $\A$,
it follows that this definition is unique so that 
there is a natural injection of $\A$ in $\Abar$. It follows that with this injection, we have
that
\be
p_l(A) = \pi_l(A) \in G_l.
\label{pil=pl}
\ee

We now show, following References \cite{aajurekproj,velhinho},  that  
$\Abar$ and $\spec$ are homeomorphic by identifying their corresponding algebras of 
continuous functions, $C(\Abar )$ and $C(\spec )$.

Let $\pol(G_l)$ denote the set of functions on $G_l$ that depend polynomially in their entries and their complex conjugates. This is the space of the functions $a_l$ of section \ref{charhba}.  Define
\be
\cyl(\Abar) := \cup_{l \in L} \; p^*_l \; \pol(G_l) \subset C(\Abar),
\ee
where $p^*_l$ is the pullback of the projections (\ref{pl}). Since the $p_l$'s are continuous, elements of $\cyl(\Abar)$ are continuous functions on $\Abar$.  An element $f \in \cyl(\Abar)$ is always of the form $f= f_l \circ p_l$ for some $l \in \L$ and some  $f_l \in \pol(G_l)$. As in section \ref{sec4} we will say that such $l$ and $f_l$ are \emph{compatible} with $f$. If $l$ is compatible with $f$ with corresponding $f_l \in \pol(G_l)$ then  $l' \geq l$ is also compatible, with corresponding function  $ f_l \circ p_{l, l'} \in \pol(G_{l'})$ (this follows from the property $p_l= p_{l,l'} \circ p_{l'}$). 
Note also that if $l$ is compatible with $f$, the surjectivity of $p_l$ implies the uniqueness of $f_l$ i.e. 
$f_l$ is the {\em only} function on $G_l$ such that $f= f_l\circ p_l$. 

 From the following four properties: i) $\cyl(\Abar)$ is a * subalgebra of $C(\Abar)$ \footnote{This follows from the fact that for given functions $f,g \in \cyl(\Abar)$, one can always find a common compatible label $l$ so that the operations of linear combinations, products and complex conjugation  in  $\cyl(\Abar)$ can  be recast  as the corresponding operations in $\pol(G_l)$. For instance: $f g= (p^*_l f_l) (p^*_l g_l)=p^*_l(f_l g_l)$.};    ii) the constant function belongs to  $\cyl(\Abar)$;  iii) $\cyl(\Abar)$ separates points in $\Abar$ (since the `coordinates' $x_l$  belong to $\cyl(\Abar)$); and iv) $\Abar$ is compact and Hausdorff, it follows from  the Stone-Weierstrass theorem  that  the completion of $\cyl(\Abar)$ in the sup norm coincides with  $C(\Abar)$ \cite{velhinho}.

We now show that $\cyl(\Abar)$ is isomorphic to  $\hba$.  Given $f \in \cyl(\Abar)$ and $l \in \L, f_l \in \pol(G_l)$ compatible with $f$, we define the map
\ba
T : \cyl(\Abar) & \to & \hba \\
 f = f_l \circ p_l& \mapsto & T(f)= f_l \circ \pi_l.\label{T}
\ea
From the properties of compatible labels and functions  described in this section for elements of $\cyl(\Abar)$ and in section \ref{sec4} for for elements of $\hba$, it follows that $T(f)$ in (\ref{T}) is independent of the choice of $l$ and that $T$ is an algebra homomorphism. 
Furthermore, by virtue of Eq. (\ref{normF}) and the surjectivity of $p_l \, \forall \,  l$, 
it follows that $||T(f)||_{\hba}=||f||_{\cyl(\Abar)}$. 


Going from  $\hba$ to $\cyl(\Abar)$,  recall from section \ref{charhba} that given $a \in \hba$, 
we can find a compatible $l \in \L$ such that $a[A]=a_l(\pi_l[A])$ with $a_l \in \pol(G_l)$.  
For any such $a_l$ we define
$f:=a_l \circ p_l$. We now show that this definition is independent of the choice of compatible $l$.
Accordingly let $l,l'$ be compatible with $a$. Consider any $l''$ compatible with $a$ with $l''\geq l,l'$.
From equation (\ref{rellabels}) we have that $a_{l}\circ p_{ll''}= a_{l'}\circ p_{l'l''}= a_{l''}$ from which it 
follows that $a_l\circ p_l = a_l\circ p_{ll''}\circ p_{l''}= a_{l'} \circ p_{l'}$.
Thus $f:=a_l \circ p_l$ defines the same element of $\cyl (\Abar )$ regardless of the choice of $l$. 
The uniqueness of $a_l$ given $a,l$ (see section 4B) then implies that this map from $\hba$ to $\cyl(\Abar)$
is injective. Finally, it can easily be verified  that 
this map is the inverse map of (\ref{T}).

It follows that $\cyl(\Abar)$ and $\hba$ are equivalent as normed, $*$ algebras. 
Hence their completions are isomorphic. By Gel'fand theory it follows that $\Abar$ and $\spec$ are homeomorphic
which completes our characterization of the quantum configuration space as a projective limit space.

As an application of this characterization, we demonstrate a curious `cartesian' structure of $\Abar$. 
Note that the Master Lemma of section \ref{sec3}  is a statement of  a certain `algebraic independence' of 
the `$U(1)$ probes' (namely the  background exponentials)  and the `$SU(2)$ probes' (namely the edge holonomies).
This suggests that the quantum configuration space may admit a split into a `holonomy' related part and a `background
exponential part'.  Indeed, a product structure of this sort is implied by the characterization of section \ref{sec4}C
wherein we showed that as a point set $\spec$ could be identified with $\hom (\P, SU(2))\times \hom (\E, U(1))$. 
However no topological information is available in this characterization.  
We would like to see if the product structure persists when $\hom (\P, SU(2))$, $\hom (\P, SU(2))$ are equipped
with suitably defined topologies so that the Gel'fand topology of $\spec$ can be realised as a product 
topology. We found it difficult to show this using $C^*$ algebraic methods because  
the norm on $\hbabar$ intertwines the properties of the holonomy and the background exponential structures. 
We now show that the projective limit characterization of $\spec$ allows an immediate demonstration
of the desired result.

Recall from section \ref{sec2C} that $\L= \Lh \times \Lb$ with $\Lh$ and $\Lb$ described in sections \ref{sec2A} and \ref{sec2B} respectively.  Each label set can be separately used to construct the projective limit spaces $\Ah$ and  $\Ab$. The relevant ingredients for $\Ah$ are the compact spaces $G_\gamma= SU(2)^n$  with $n$ the number of edges in $\gamma$, and the maps  $p_{\gamma \gamma'}$  as described after Eq. (\ref{hepp}) (see also Appendix \ref{projlimapp}). Similarly the relevant ingredients for  $\Ab$ are the spaces  $G_\Upsilon:=U(1)^N$ with $N$  the number of electric fields  in $\Upsilon$ and   corresponding projections $p_{\Upsilon, \Upsilon'}$ as described after Eq. (\ref{Ptilde}).  Let $p_\gamma: \Ah \to G_\gamma$ and $p_\Upsilon: \Ab \to G_\Upsilon$ be the canonical projections analogous to (\ref{pl}). 
We now demonstrate that 
$\Abar$ and $\Ah \times \Ab$ are homeomorphic.

We first construct a bijection between the two spaces.
Given $\{x_l \} \in \Abar$, each $x_l \in G_l$ is given by a pair $x_\gamma \in G_\gamma$ and $x_\Upsilon \in G_\Upsilon$ where $l=(\gamma,\Upsilon)$ so that $G_l=G_\gamma \times G_\Upsilon$. The consistency condition on the $x_l$'s implies consistency of the $x_\gamma$'s and $x_\Upsilon$'s  so that $\{x_\gamma \}$ defines an element in $\Ah$ and $\{x_\Upsilon \}$ an element in $\Ab$. 
Conversely, given $\{x_\gamma \} \in \Ah$ and $\{x_\Upsilon \} \in \Ab$ the corresponding point in $\Abar$ is given by $x_l \equiv x_{(\gamma,\Upsilon) }:=(p_\gamma(\{ x_{\gamma'} \}), p_\Upsilon(\{x_{\Upsilon'} \}) \in G_l $.  $\{x_l \}$ so defined satisfies the consistency conditions and hence defines an element of $\Abar$ which corresponds to the inverse of the previous mapping. 

We now show that this bijection between $\Abar$ and $\Ah \times \Ab$ is a homeomorphism. Recall that the 
topologies of  $\Abar$, $\Ah$,  $\Ab$ are generated by inverse projections $p_l^{-1}, p_{\gamma}^{-1},p_{\Upsilon}^{-1}$ 
of open sets in $G_l, G_{\gamma}, G_{\Upsilon}$. Given $l=(\gamma, \Upsilon)$ and open sets 
$U_{\gamma} \subset G_{\gamma}, U_{\Upsilon} \subset G_{\Upsilon}$, the bijection  between 
$\Abar$ and $\Ah \times \Ab$ 
identifies the open set $p_{l}^{-1} (U_{\gamma}\times U_{\Upsilon}) \in \Abar $ 
with the open set $p_{\gamma}^{-1} (U_{\gamma}) \times p_{\Upsilon}^{-1} (U_{\Upsilon})\in \Ah \times \Ab$.
Since the product topology on $G_l$ is generated by rectangle sets and since
\be
p_l^{-1} (\cup_{\bf \alpha} (U^{\bf \alpha}_{\gamma} \times U^{\bf \alpha}_{\Upsilon})) 
= \cup_{\bf \alpha}p_l^{-1}(U^{\bf \alpha}_{\gamma} \times U^{\bf \alpha}_{\Upsilon}), \;\;\;
p_l^{-1} (\cap_{i} (U^{i}_{\gamma} \times U^{i}_{\Upsilon})) 
= \cap_{i}p_l^{-1}(U^{i}_{\gamma} \times U^{i}_{\Upsilon}), \;\;\;
\ee
for some (possibly non-denumerable) label set $\bf\alpha$ and finite label set $i$, it follows that 
the bijection is indeed a homeomorphism thus completing the proof.

To relate this result with that of section \ref{sec4}, let us denote by $\spech$  the spectrum of  $\habar$ and $\specb$ that of $\Babar$. The arguments of the present and previous sections may be reproduced for each of the algebras $\Ba$ and $\ha$ to conclude that $\Ab \simeq \specb \simeq \hom(\E,U(1))$
\footnote{In order to show this we use the fact that, just as for $\ha$, elements of $\Ba$ seperate points in $\A$.}
 and  $\Ah \simeq \spech \simeq \hom(\P,SU(2))$ (the latter being the standard characterizations in LQG). Thus, the product structure presented here coincides with that of section \ref{sec4}. 

\subsection{Measure theoretic aspects of the projective limit space}\label{sec5B}
Recall from Appendix \ref{projlimapp} that we have two equivalent projective limit constructions of $\Abar$. The first, which 
we have used hitherto in this section, is based on the preordered directed set of labels $\L = \{l\}= \{(\gamma, \Upsilon )\}$. 
The second is based on the partially orderered directed
label set ${\hat{\L}}= \{{\hat l}\}= \{({\hat \gamma}, {\hat \Upsilon} )\}$ where, as detailed in Appendix \ref{projlimapp},
${\hat \gamma}$ corresponds to the path subgroupoid  of $\P$ generated by the edges in $\gamma$ and ${\hat \Upsilon}$
to the abelian subgroup of $\E$ generated by the electric fields in $\Upsilon$.
In what follows we shall follow the 
argumentation of Reference  \cite{velhinho}. Since this reference uses a partially ordered label set 
in its analysis, we shall use the second, partially ordered directed label set, characterization of $\Abar$.

Let the Haar measure on $G_{\hat l}$ be $\mu_{\hat l}$. Note that the  Haar measure $\mu_{\hat l}$ 
is a regular Borel measure.
As shown in  Appendix \ref{plimapp3}, the set of measures
$\{ \mu_{\hat l} \}$ satisfy  the consistency condition   $(p_{{\hat l},{\hat l}'})_* \mu_{{\hat l}'}= \mu_{\hat l}$ 
whenever ${\hat l}' \geq {\hat l}$, where $(p_{{\hat l},{\hat l}'})_* \mu_{{\hat l}'}$ is the push-forward measure.
Let $C( \Abar) \equiv C(\spec )\equiv \hbabar$ be the $C^*$ algebra of continuous functions on $\Abar$.  
Our demonstration above that $\cyl (\Abar ) = \hba$ implies that 
$\cyl (\Abar)$ is  dense in $C( \Abar)$. It then follows from Proposition 4,
section 3.2 of \cite{velhinho} that the consistent family of (regular Borel) Haar measures $\{\mu_{\hat l}\}$ 
defines a unique regular Borel measure on $\Abar$. A natural question is if this projective limit measure
is the same as the measure $\muks$.\footnote{
Note that the identification of $\Abar$ with $\spec$ as a topological space implies the identification of 
corresponding Borel algebras so that this question is well posed.}
We now show that the answer to this question is in the affirmative.
The proof of Proposition 4 in Reference \cite{velhinho} uses the 
fact that the consistent set of measures define a positive linear function (PLF) on $C( \Abar)$. From Appendices \ref{plfintapp}
and \ref{plimapp3},
it follows that this PLF is exactly the KS PLF of equation (\ref{defplf}). It immediately follows from this fact,
together with 
the unique association of the measure $\muks$ with the KS PLF via the Riesz-Markov theorem,
 that the projective limit measure is $\muks$.

As an application of the projective limit characterization of $\muks$, 
it is straightforward 
to check that a simple adaptation of the proof of Marolf and Mour\~ao \cite{josedon} for the LQG case shows that,
as in LQG, the classical configuration space $\A$ lies in  a set of measure zero in the quantum configuration space
$\Abar$. The main feautures  of this adaptation are as follows
(we assume familiarity with the notation and contents
of Reference \cite{josedon}):\\
\noindent (a)Choose the subsets  $\triangle^{ \{\epsilon_i\}_{i=1}^n    }$    of $SU(2)^n$ as in Reference \cite{josedon}.
For each  $q \in (0,q_0]$ (for some fixed $q_0, 0< q_0 <1$), choose 
$\{\epsilon_i\}_{i=1}^{\infty}= \{\epsilon_i^{(q)}\}_{i=1}^{\infty}$ as in \cite{josedon}.\\
\noindent (b) For each fixed $q$, replace the set of shrinking (independent) hoops $\{\beta_i\}$ of \cite{josedon} 
by a set of shrinking edges $\{e_i\}$. To specify this
edge set, fix a coordinate chart $\{x,y,z\}$ around some point $p_0 \in \Sigma$ so that $p_0$ is at the origin
and let $e_i$ be the straight line along the $z$-axis from the origin  to $(0,0,\epsilon^{(q)}_i\delta_i)$ where
$\delta_i$ is as defined in \cite{josedon}.\\
\noindent (c)  Use the small edge expansion for the edge holonomy of a connection $A\in \A$ to show that its 
edge holonomy is confined to a neighbourhood of the identity of size $\epsilon_i^{(q)}$ for sufficiently large $i$. \\
It is then easy to see that the desired result follows by a repetition of  the proof of  Marolf and Mour\~ao 
(without the complications of quotienting 
by the action of gauge transformations, since we are interested in $\A$ rather than $\A/\G$).

We leave
other applications of  projective techniques  (such as the definition of a host of projectively consistent 
`differential geometric' structures \cite{aajurekproj,ttbook})  to future work.

\section{The holonomy-background exponential-flux algebra} \label{sec6}
Our construction of the holonomy-background exponential-flux algebra parallels that of the 
holonomy-flux algebra in References \cite{acz,lost}
and we assume familiarity with those works.
The only minor difference between our treatment and theirs is that we restrict attention to 
polynomial cylindrical functions of the form (\ref{psi}) whereas the cylindrical functions 
of \cite{acz,lost} comprise all continuous functions rather than only polynomials.
In section \ref{sec6A} we construct the holonomy-background exponential-flux algebra. In section \ref{sec6B}
we use an identity of 
Stottmeister and
Thiemann \cite{ttks} to illustrate the difference between the holonomy-background exponential-flux algebra
and the LQG holonomy flux algebra.
In section \ref{sec6C} we show that the KS representation is a representation of the holonomy-background exponential-flux
algebra.

\subsection{Construction of the classical and quantum  algebras.} \label{sec6A}
Let $a \in \hba$ and let $l\in \L$ be compatible with $a$ (see section \ref{sec4}).  Then it is straightforward to
verify  that:
\ba
\{ a[A], F_{S,f}\}& =: & X_{S,f}a[A]
\label{eax}\\
X_{S,f} & :=& \Xh_{S,f} + \Xb_{S,f} \label{x}\\
\Xh_{S,f} &:= & \sum_{e_i\in l}(\Xh_{S,f}h_{e^iC_i}^{\quad D^i}) \frac{\partial}{\partial h_{e^iC_i}^{\quad D^i}} 
\label{xs} \\
\Xb_{S,f} &:=&  \sum_{E_I\in l}(\Xb_{S,f}\beta_{E_I}) \frac{\partial}{\partial \beta_{E_I}} .
\label{xu}
\ea
Here $(\Xh_{S,f}h_{e_i})$ is defined exactly as in LQG so that its evaluation involves the 
appropriate action of $SU(2)$ invariant vector fields on $h_{e_i} \in SU(2)$.                 
The evaluation of  $(\Xb_{S,f} \beta_{E_I})$ involves the 
analogous action of the  $U(1)$ invariant vector field  on $\beta_{E_I}\in U(1)$:
\be
\Xb_{S,f}\beta_{E_I} = 
\int_{S}dS_a f^i E^a_{iI} \frac{\partial}{\partial \theta}e^{i\theta}\vert_{e^{i\theta}= \beta_{E_I}} 
                       =  i F_{S,f}(E_I) \beta_{E_I} .
\label{xu1}
\ee
From equations (\ref{x})-(\ref{xu1}) it follows that the operators $X_{S,f},\Xh_{S,f},\Xb_{S,f}$ all act
as {\em derivations} on $\hba $ i.e. they map $\hba$ into itself and their actions obey the 
Leibniz rule (see equation (\ref{deriv}) below). It is also easy to check that 
\be
[\Xb_{S,f},\Xb_{S',f'}]= [\Xb_{S,f},\Xh_{S',f'}] =0\;\;\; {\rm on} \;\; \hba .
\label{commute}
\ee
Next, note that if we choose $a \in \hba$ such that it depends only on the holonomies, $X_{S,f}$ acts exactly as in 
LQG. In other words, the action of $X_{S,f}$ restricted to $\ha \subset \hba$ ($\ha$ is defined in \ref{sec4C})
 is exactly the LQG action.
It then follows, similar to the LQG case, that while the commutator of a pair of derivations $[X_{S,f},X_{S',f'}]$ on 
$\hba$ is itself, in general, not of the form $X_{S'',f''}$, this commutator still acts as a derivation on $\hba$.
As in the LQG case, consider, the finite span of objects $\Vder$ of the form:
\be
a X_{S,f}, a [X_{S_1,f_1},X_{S_2,f_2}], a [ X_{S_1,f_1},[...[X_{S_{n-1},f_{n-1}}, X_{S_n,f_n}]..]]
\label{vderiv}
\ee
(Note that the classical correspondents of these objects are $a F_{S,f}, 
a \{F_{S_1,f_1},F_{S_2,f_2}\}, a \{ F_{S_1,f_1},\{...\{F_{S_{n-1},f_{n-1}}, F_{S_n,f_n}\}..\}\}$,
where $ a\in \hba$).
It follows that every $Y\in \Vder$ acts as a derivation on $\hba$ i.e.
\be
Y:\hba \rightarrow \hba, \;\;\;Y(ab) = Y(a)b + a Y(b), \;\; \forall a,b\in \hba .
\label{deriv}
\ee
Next, define the vector space $\mathfrak{A} = \hba  \times \Vder$. We define the  $*$ operation on $\mathfrak{A}$ by:
\be
(a,Y)^* = ({\bar a}, {\bar Y}) \;\;\; {\rm where} \;\;\; {\bar Y}(b):= \overline{Y(\bar{b})}
\ee
where $\bar{a}$ denotes the complex conjugate of $a$.
From the definitions (\ref{x})-(\ref{xu1}), it follows that ${\bar X_{S,f}}= X_{S,f}$. This can then be used to
show that multiple commutators of the $X_{S,f}$'s are also invariant under the ` ${\bar{\;}}$ ' relation. It then follows 
straightforwardly that the $*$ operation maps $\mathfrak{A}$ to itself and defines a $*$ relation on $\mathfrak{A}$.
Finally, we define the Lie bracket ${\bf [},{\bf]}$ on $\mathfrak{A}$ by
\be
{\bf [} (a,Y), (a'Y'){\bf ]} = ( Y'(a) - Y(a'), [Y,Y'])
\label{lieu}
\ee
where $[Y,Y']\in \Vder$ is the commutator of the derivations $Y, Y'$ on $\hba$. 
We refer to
$(\mathfrak{A}, {\bf [},{\bf ]})$ as the {\em classical holonomy-background exponential-flux algebra}.
It is the exact counterpart of the (classical) ACZ holonomy-flux algebra
(referred to as $(\mathfrak{A}_{\text{class}}, \{, \})$ in Reference \cite{lost}) underlying the standard LQG representation.
The classical Lie algebra $\mathfrak{A}$ can be converted to its quantum counterpart $\hat{\mathfrak{A}}$ through the steps
of section 2.5 of Reference \cite{lost}.

In the next section we comment on the differences between the algebra $\mathfrak{A}$ and its LQG counterpart
$\mathfrak{A}_{\text{class}}$ \cite{lost}.
For notational convenience, 
 we shall denote $\mathfrak{A}_{\text{class}}$
by $\mathfrak{A}^{\lqg}$.

\subsection{On the difference between $\mathfrak{A}$ and $\mathfrak{A}^{\lqg}$} \label{sec6B}

As is apparent from the previous section, the construction of $\mathfrak{A}$ differs from that of 
$\mathfrak{A}^{\lqg}$ due to the added structure provided by the background exponentials.
More in detail the generators of $\mathfrak{A}$ (see equation (\ref{vderiv}))  differ from those of $\mathfrak{A}^{\lqg}$
(equation (19) in Reference \cite{lost})
in two ways. First, the cylindrical function $a$ in (\ref{vderiv}) depends on the background exponentials as well as the holonomies
so that $a \in \hba$ whereas the cylindrical function $\Psi$ in (19) of \cite{lost} depends only on the holonomies
so that $\Psi \in \ha$. The second difference is that the derivation $X_{S,f}$ and its commutators in 
(\ref{vderiv})  inherit their algebraic properties from their realization as derivations on $\hba$ whereas
the corresponding LQG objects in (19) of \cite{lost} inherit theirs  from their realization as derivations
on $\ha$. In contrast to the first, the second difference is a bit subtle. 
To see it explicitly, we turn our attention to the beautiful example considered by Stottmeister and Thiemann in Reference 
\cite{ttks}.

Accordingly, consider $(0, [X_{S,f_1}[X_{S,f_2},X_{S,f_3}]]) \in \mathfrak{A}^{\lqg}$.
From the  Stottmeister-Thiemann identity \cite{ttks}, we have that 
\be
[X_{S,f_1}[X_{S,f_2},X_{S,f_3}]]= \frac{1}{4} X_{S,[f_1,[f_2,f_3]]}
\label{stid}
\ee
where both the left hand side and the right hand side are derivations on $\ha$.
This implies  the identification of the elements $(0, [X_{S,f_1}[X_{S,f_2},X_{S,f_3}]])$
and $(0, \frac{1}{4} X_{S,[f_1,[f_2,f_3]]})$ in the algebra $\mathfrak{A}^{\lqg}$.

Let us now consider $(0, [X_{S,f_1}[X_{S,f_2},X_{S,f_3}]])$ as an element of $\mathfrak{A}$.
From equations (\ref{x}) and (\ref{commute}) it follows that 
\be
(0, [X_{S,f_1}[X_{S,f_2},X_{S,f_3}]])= (0, [\Xh_{S,f_1}[\Xh_{S,f_2},\Xh_{S,f_3}]] )
\label{x,x,x}
\ee
From (\ref{stid}), (\ref{xs})  it immediately follows that 
\be
[\Xh_{S,f_1}[\Xh_{S,f_2},\Xh_{S,f_3}]] = \frac{1}{4} \Xh_{S,[f_1,[f_2,f_3]]} .
\label{stids}
\ee
Now, from (\ref{x}) it follows that $\Xh_{S,[f_1,[f_2,f_3]]} \neq X_{S,[f_1,[f_2,f_3]]}$ because of the missing
`$U(1)$' contribution, $\Xb_{S,[f_1,[f_2,f_3]]}$, which in turn means that in contrast to the LQG case,
the two elements 
$(0, [X_{S,f_1}[X_{S,f_2},X_{S,f_3}]]_\lqg)$
and $(0, \frac{1}{4} X_{S,[f_1,[f_2,f_3]]})$ are {\em not} identified in the algebra $\mathfrak{A}$.

\subsection{The KS representation and $\hat{\mathfrak{A}}$} \label{sec6C}
As shown in Reference \cite{lost} elements in $\hat{\mathfrak{A}}$ are of the form 
${\hat a}{\hat X}_{S_1,f_1}{\hat X}_{S_2,f_2}..{\hat X}_{S_n,f_n}$ where $\hat a$ and ${\hat X}_{S_i,f_i}$ are the
quantum correspondents of $a\in \hba$ and $F_{S_i,f_i}$. The algebraic properties of $\hat a$ derive from those
of $a \in \hba$ and the algebraic properties of ${\hat X}_{S_i,f_i}$ derive from the algebraic properties
of $X_{S_i,f_i}$ as derivations on $\hba$. Thus the algebraic properties of elements in $\hat{\mathfrak{A}}$ 
derive from those of $\hba$ and derivations thereon.

Next we note the following:\\
\noindent (1) There is an algebraic isomorphism between cylindrical
functions on $\spec$, $\cyl(\spec)$,  and $\hba$ (see section \ref{sec4}).\\
\noindent (2) The KS representation is an $L^2(\spec, d\muks)$ representation. 
The space of cylindrical functions $\cyl(\spec)$ is dense in $L^2(\spec, d\muks)$.
The operators  
${\hat a}, {\hat F}_{S,f}$ map $\cyl (\spec)$ into itself and their actions can be inferred from the 
algebraic isomorphism between $\cyl (\spec )$ and $\hba$ (see section \ref{sec4E}).
In particular given $\Psi \in \cyl(\spec)$ with correspondent $\Psi[A]\in \hba $ :\\
\noindent (i) ${\hat a}$ acts by multiplication so that ${\hat a}\Psi: = a \Psi$,
where $a\in \cyl(\spec)$ is the correspondent of the element $a[A] \in \hba$.\\
\noindent (ii) From section \ref{sec4E} and equation (\ref{eax}), the correspondent of ${\hat F}_{S,f}\Psi$ in $\hba$ is $- i X_{S,f}\Psi[A]$ so 
that the  algebraic properties of ${\hat F}_{S,f}$ are determined by those of the derivation  $X_{S,f}$ on $\hba$.\\
\noindent (3) The KS inner product  implements the $*$ relations on
${\hat a}, {\hat F}_{S,f} \in \hat{\mathfrak{A}}$ 
 as adjointness relations.\\

The discussion in the first paragraph of this section together with (1)-(3) above show that the 
KS representation is indeed a representation of $\hat{\mathfrak{A}}$.

\section{Discussion} \label{sec7}

In this work we have shown that the KS representation admits structural characterizations which are the counterparts of 
LQG ones. An immediate question is whether these characterizations can also be used to show a LOST-Fleischhack
type \cite{lost,christian} uniqueness theorem based on the holonomy--background exponential--flux algebra of 
section \ref{sec6}. If so, the KS representation would then be the unique representation of this algebra with a 
cyclic, diffeomorphism
invariant and $SU(2)$ gauge invariant state and at the kinematic level, there would be little to choose
between  the KS and the LQG representations. A key question is then if any progress is possible in the KS 
representation with regard to the quantum dynamics. While References \cite{hs,miguelme} suggest that there is no
fundamental obstruction to the imposition of 
$SU(2)$ gauge invariance and spatial diffeomorphism invariance, we do not know if the Hamiltonian constraint
can be defined in this representation. Of course, if one takes the view that the KS representation 
is some sort of effective description for smooth spatial geometry, and that the underlying fundamental description is
that of LQG, this question is moot.

The results presented in this work and summarised in detail in section \ref{sec1} have been derived in the context 
of compact (without boundary) spatial topology. 
However, our main interest in the KS representation is in its possible application to asymptotically flat quantum gravity.\footnote{For an application to the case of parametrized field theory,  see Refs. \cite{sandipan1,sandipan2}.}
In contrast to the compact case, in the asymptotically flat case the classical connection and its conjugate triad field 
are required to satisfy detailed boundary conditions at spatial infinity.

The triad field is required to approach a fixed
flat triad at spatial infinity. It is difficult to tackle the triad  boundary conditions in an LQG like representation
because of the contrast of the discrete spatial geometry underlying LQG with the  smooth, 
asymptotically flat spatial geometry
in the vicinity of spatial infinity. In particular, the spatial geometry needs to be excited in a non-compact
region which means that the LQG spin networks of the compact case need to be generalized so 
as to have infinitely many edges and vertices \cite{itp}. Moreover a suitably coarse grained view of the 
quantum  spatial geometry in the vicinity of spatial infinity must coincide with an  asymptotically flat one. 
On the other hand, the KS representation offers the possibility of already accounting for 
smooth spatial geometry without coarse graining, and, asymptotically flat spatial geometry
 without the consideration of KS spinnet graphs
with infinitely many edges \cite{miguelme}; in brief this may be achieved by restricting attention to KS spin net states
whose graphs have a finite number of compactly supported edges but whose triad label satisfies the asymptotic conditions.
The boundary conditions on the connection are, however, much harder to tackle. This is so because
the natural spinnet basis has a much more controllable behaviour with respect to the electric flux
operators both in the LQG and the KS representations.  Since we already have a possibility of encoding 
the triad boundary conditions in the KS representation, let us focus on the issue of connection 
boundary conditions in the KS (as opposed to the LQG) representation. 

As we have shown, in the compact case, one way in which the quantum configuration space is tied to the classical
configuration space is that the latter embeds into the former as a dense set. One may hope that something similar happens
in the asymptotically flat case, namely, the quantum configuration space is the topological completion of the 
space of semianalytic connections {\em satisfying the asymptotic conditions}. While this would be one way in which 
the classical boundary conditions on the connections leave their imprint in the quantum theory, one may worry that
since the quantum configuration space is larger than the classical one, perhaps a (for example, measure theoretically,) 
large number of quantum connections
could be thought of as violating these boundary conditions. We now argue that this fear could be misplaced.
In the holonomy-background exponential algebra, $\hba$,
classical connections are integrated against one dimensional edges to give holonomies and against three dimensional
background fields to give background exponentials.
If, as mentioned in the previous paragraph, in the asymptotically flat case
we restrict the edges of interest to be confined to compact regions, it is only the background fields
which sense the asymptotic behaviour of the connection. Let us then focus only on the structures associated with the
background exponentials. 
A preliminary analysis indicates that the 
the fields which label the background exponential functions  satisfy 
 the ``maximally'' permitted asymptotic 
behaviour  which allows their integrals with respect to (classical) connections to be well defined.
If it transpires that the quantum connections, as in the compact case, define homomorphisms from  
the abelian group of fields, subject to this `maximal' asymptotic behaviour, the very existence of these homomorphisms
could be reasonably interpreted as the imposition of the classical asymptotic behaviour on the quantum 
configuration space.

Thus, the KS representation offers hope that the complications arising due to asymptotic flatness
are not insurmountable, at least at the level of quantum kinematics. 
We are at present engaged in working out the ideas sketched above. To conclude, we remark that if
these efforts meet with success, they may also shed light on how to generalise LQG to  asymptotically
flat spacetimes  so as to retain the  most remarkable feature of the theory, namely the fundamental discreteness of space.
\\

\textbf{Acknowledgements:} 
 We thank Alok Laddha for valuable discussions and for his help in the projective limit construction, including his identification of  electric fields as appropriate labels therein. We thank Hanno Sahlmann for correspondence in the early stages of this work, for valuable conversations and for his constant encouragement. 
We thank Thomas Thiemann for taking the time to help us with our queries regarding Reference \cite{ttks} and for
valuable comments. We thank
Jos\'e Velhinho for his invaluable help with regard to our questions on projective structures and Bohr compactifications. 
We are very grateful to Fernando Barbero, HS, TT  and Eduardo Villase\~nor for going through a draft version of this work. MC is deeply grateful to AL for helping him learn most of the  LQG-related topics that were needed for the present work.

\appendix
\section{Proof of the Master Lemma} \label{applemma}
\noindent  Step (i): \\
The $N$ background electric fields $E_i$ are rationally independent but not necessarily linearly independent. Let $m \leq N$ be the dimension of the linear span of the background electric fields and assume $\{E_1,\ldots,E_N\}$ is ordered so that the first $m$ electric fields are linearly independent. The last $p:=N-m$ electric fields can then be written as linear combinations of the first $m$ ones:
\be
E_{m+j}= \sum_{\mu=1}^{m} k^\mu_j E_\mu , \quad j=1,\ldots,p , \label{Emj}
\ee
for some real constants $k^\mu_j$. Next,  let $A^\nu, \nu=1,\ldots,m$, be $m$ $su(2)$-valued one-forms  satisfying:\footnote{To explicitly obtain $m$ $su(2)$-valued one-forms $A^\nu$ satisfying (\ref{dualE}), we introduce a semianalytic metric $h_{ab}$ on $\Sigma$ which defines an inner product on the space of background electric fields by $\bra E, E' \ket := \int h^{-1/2} h_{ab}\tr[E^a E'^b]$. Since $E_1,\ldots,E_m$ are linearly independent, and $\bra, \ket$ positive definite, the $m \times m$ matrix  $\bra E_\mu , E_\nu \ket , \mu, \nu =1,\ldots,m$ is invertible. Let $c_{\mu \nu}$ be its inverse, so that $\sum_{\rho=1}^{m} \bra E_\mu , E_\rho \ket c_{\rho \nu} = \delta_{\mu \nu}$.  It is then easy to verify that the one-forms $A^{\nu}_a :=  h^{-1/2} \sum_{\rho=1}^{m}  c_{\rho \nu} h_{ab} E^b_\rho$  satisfy (\ref{dualE}).
}
\be
\int \tr[E_\mu^a  A^{\nu}_a]= \delta_\mu^\nu ,\quad \mu,\nu = 1,\ldots,m , \label{dualE}
\ee
 and consider the  $m$ parameter family of one-forms:
\be
A_{\vec{t}}:=\sum_{\mu=1}^{m} t_\mu A^\mu. \label{Avect}
\ee
The $U(1)^N$ part of the map (\ref{phiA})  restricted to the $m$ parameter family of connections (\ref{Avect}) induces the following map from $\reals^m$ to $U(1)^N$:
\be
(t_1,\ldots,t_m) \mapsto (e^{i t_1}, \ldots, e^{i t_m},e^{ i t_\mu k^\mu_1}, \ldots, e^{ i t_\mu k^\mu_p}).  \label{RmtoU1N}
\ee
Our aim is to use the map (\ref{RmtoU1N}) to reproduce with arbitrary precision the given $N$ phases $(e^{i \theta_1},\ldots,e^{i \theta_N}) \in U(1)^N$. The first $m$ phases can be exactly reproduced by taking:
\be
t_\mu = \theta_\mu + 2 \pi n_\mu, \quad n_\mu \in \integers, \;  \mu=1,\ldots,m .
\ee
We are then left with the $m$ integers $\{ n_\mu \}$ to approximate $p$ phases, the relevant map being: 
\be
(n_1,\ldots,n_m) \mapsto  (e^{ i 2 \pi n_\mu k^\mu_1}, \ldots, e^{ i 2 \pi n_\mu k^\mu_p}). \label{ZmtoU1N}
\ee
Now, the condition of rationally independence of the  $N$ electric fields translates into the following condition of rational independence of $N$ vectors in $\reals^{m}$: \emph{The canonical basis  $\vec{e}_i \in \reals^{m}, i =1,\ldots,m$ (with components $(\vec{e}_i)^\mu=\delta^\mu_i$), together with the  vectors $\vec{k}_j \in \reals^{m}, j =1,\ldots,p$ (with components $(\vec{k}_i)^\mu=k^\mu_j$), are rationally independent.} The example (26.19 (e)) of Ref. \cite{hewittross} shows that, under this condition, the range of the map (\ref{ZmtoU1N}) is dense in $U(1)^p$.\footnote{This results also follows from  theorem IV in section III.5 of \cite{cassels}.} This implies that given $\delta>0$, we can find $\vec{t}^{(\delta)} \in \reals^m$ such that $e^{i t^{(\delta)}_\mu } = e^{i \theta_\mu},  \mu=1,..,m$ and $|e^{i t^{(\delta)}_\mu k^\mu_j }- e^{i \theta_{m+j}}|< \delta, j=1,\ldots,p=N-m$.  Setting  ${\bar A}^{B, \delta}:= A_{\vec{t}^{(\delta)} }$  we obtain the desired connection satisfying  (\ref{budelta}).\\

\noindent Step (ii):

Let 
$p_\alpha$ be a point on the open edge $\tilde{e}_\alpha - \{b(e_\alpha),f(e_\alpha)\}$.
Since $\Sigma$ is Hausdorff, there exists an open neighbourhood $U_{\alpha}$ of $p_{\alpha}$ such that 
$U_{\alpha}$ seperates $p_{\alpha}$ from the points $b(e_\alpha),f(e_\alpha)$. Further, $U_{\alpha}$ can be chosen
such that $U_\alpha \cap \tilde{e}_\beta = \emptyset$ for $\alpha \neq \beta$; else 
$p_{\alpha}$ is an accumulation point of a sequence in $\tilde{e}_\beta$, which, by virtue of the compactness of 
$\tilde{e}_{\beta}$ implies that 
 $p_\alpha \in \tilde{e}_\beta \cap \tilde{e}_\alpha$, contradicting the condition that $\tilde{e}_\alpha$ and $\tilde{e}_\beta$ can only intersect at 
their endpoints. A similar argument implies that $U_{\alpha}, \alpha =1,..,n$ can be chosen such that 
$U_\alpha \cap U_\beta = \emptyset$ if $\alpha \neq \beta$.

Finally, since $\tilde{e}_{\alpha}$ is a semianalytic manifold, it follows (see for example Definition A.12 of Reference
\cite{lost}) that $U_{\alpha}$ can be chosen to be small enough that it is in the domain of a
single  semianalytic chart $\chi_{\alpha}$ in which it takes the form of a ball of size $\tau$ 
 within which $\tilde{e}_{\alpha}\cap U_{\alpha}$ is 
connected and runs along a coordinate axis. Thus, we have that 
$\chi_\alpha(U_\alpha) \subset \reals^3$ with 
$\chi_\alpha(U_{\alpha} )=\{ \vec{x} \in \reals^3 |\; || \vec{x}|| < \tau \}$,  and that 
 $\chi_\alpha(U_{\alpha} \cap \tilde{e}_\alpha)=( (-\tau,\tau),0,0)$.

In the $\chi_{\alpha}$ coordinate chart, we denote balls of coordinate size $\delta$ 
centred at the origin by $B_{\alpha}(\delta )$. Accordingly we 
denote the above choice of 
$U_{\alpha}$  by $B_{\alpha}( \tau )$.  Clearly by taking $\epsilon \ll \tau$ we have $B_\alpha(2 \epsilon) \subset U_\alpha$.
\\

\noindent Step (iii):

We need to specify $f_\epsilon$ in  $B_{\alpha}( 2\epsilon )- B_{\alpha}( \epsilon )$ such that $f_\epsilon$ is semianalytic. 
Consider the  polynomial in $\reals$ given by $g(y):= c \int_{0}^y (y'(1-y'))^K d y'$, with   $c$ chosen so that $g(1)=1$ and $K > k$. Then $g$ interpolates between the constant $0$ function for $y<0$ and the constant $1$ function for $y>1$ in a $C^{K}$ manner. Setting
\be
(f_\epsilon \circ \chi_\alpha)(\vec{x})= \left\{
\begin{array}{cll}
0 & \text{for} & ||\vec{x}|| < \epsilon  \\
1  & \text{for}  & ||\vec{x}|| \geq 2 \epsilon \\
 g(\frac{1}{3}(||x||^2/\epsilon^2 -1))  &\text{for} & \epsilon \leq ||\vec{x}|| < 2 \epsilon 
\end{array} \right. 
\ee
does the job.  \\

\noindent Step (iv):

We take $A^{\epsilon}$ with support on $U_\alpha$ given by:
\be
A^{\epsilon}|_{U_\alpha}= w_\alpha \chi_\alpha^*( a_\epsilon  dx). \label{Aeps}
\ee
Here $w_\alpha \in su(2)$ are constant (but $\epsilon$ and $\delta$ dependent) $su(2)$ elements satisfying 
\be
e^{w_\alpha}= (g^1_{\alpha})^{-1} g_\alpha (g^2_{\alpha})^{-1}, \quad \alpha=1,\ldots,n, \label{walpha}
\ee
and taken to be in the ball of radius $4\pi$ of $su(2)$ that maps onto $SU(2)$ under the exponential map.   

$a_\epsilon: \chi_\alpha(U_\alpha) \to \reals$ is taken to be:
\be
a_\epsilon(\vec{x})= \left\{
\begin{array}{cll}
c' (\epsilon^2-||\vec{x}||^2)^K & \text{for} & ||\vec{x}|| \leq \epsilon  \\
0  & \text{for}  & ||\vec{x}|| >  \epsilon \label{defaeps}
\end{array} \right. 
\ee
with $K >k$ and $c'$ chosen so that $\int_{-\epsilon}^{\epsilon} a_{\epsilon}(x,0,0) dx=1$. This last condition, together with the choice of $w_\alpha$ (\ref{walpha}) guarantees the required condition (\ref{step4}). 
\\
\noindent Step (v):

$A^\epsilon$ was constructed so that the holonomies of  $A^{\delta}:= f_{\epsilon} {\bar A}^{B,\delta}+ A^{\epsilon}$ along the edges $e_\alpha$ exactly reproduce the group elements $g_\alpha$ and hence (\ref{hgdelta}) is satisfied with $C_1=0$. For the $U(1)$ elements we have
\ba
\left| e^{i\int E_I \cdot A^{\delta}} - e^{i \theta_I } \right| & =& \left| e^{i \int E_I \cdot {\bar A}^{B,\delta} }e^{-i \theta_I} -  e^{i \int[ (1-f_\epsilon) E_I \cdot  {\bar A}^{B,\delta}-E_I \cdot A^\epsilon]} \right| \\
& \leq & \left| e^{i \int E_I \cdot {\bar A}^{B,\delta} }e^{-i \theta_I} - 1 \right| +  \left| e^{i \int[ (1-f_\epsilon) E_I \cdot  {\bar A}^{B,\delta}-E_I \cdot A^\epsilon]}  -1 \right| \label{phb}
\ea 
From step (i) above, the first term in \ref{phb} is bounded by $\delta$. The phases in the second term can be bounded by:
\ba
\left| \int (1-f_\epsilon) E_I \cdot  {\bar A}^{B,\delta} \right| & \leq & \sum_{\alpha=1}^n \left| \int_{B_\alpha(2\epsilon)} (1-f_\epsilon)  E_I \cdot  A^{(\beta)}_\delta \right|\\
& \leq & c_1(\delta) \epsilon^{3}
\ea
for some constant $c_1(\delta)$, and
\ba
\left| \int E_I \cdot A^\epsilon \right| & \leq & \sum_{i=1}^n \left| \int_{||\vec{x}|| < \epsilon} \tr[((\chi_i)_* E_{I})^x w_i] a_\epsilon  dx dy dz \right| \\
& \leq & c_2(\delta) \epsilon^2 \label{2ndepsterm}
\ea
for some constant $c_2(\delta)$. Here we used the fact that $\tr[((\chi_i)_* E_{I})^x w_i]$ has some $\epsilon$ independent bound and that $\int |a_\epsilon| dx dy dz$ has an order $\epsilon^2$ bound as follows from the condition on $c'$ described after Eq. (\ref{defaeps}) . By Taylor expanding, we conclude that, for given $\delta$ and sufficiently small $\epsilon$, the second term in (\ref{phb}) has an $\epsilon^2$ bound:
\be
\left| e^{i \int[ (1-f_\epsilon) E_I \cdot  {\bar A}^{B,\delta}-E_I \cdot A^\epsilon]}  -1 \right| < c(\delta) \epsilon^2
\ee
for some constant $c(\delta)$. Thus, if for given $\delta$ we chose $\epsilon$ such that
\be
\epsilon \ll (\delta/c(\delta))^{1/2}
\ee
we achieve the desired  bound (\ref{budelta}) with $C_2 = 2$.

\section{Assorted proofs} 
\subsection{Generating set (\ref{genset}) }\label{gensetapp}
If the  $M$ electric fields $E'_1, \ldots, E'_M$, are algebraically independent, then $M=N$. If not, then there exists $M$ integers $q_{i}, i=1,\ldots,M$, not all of them zero, such that $\sum_{i} q_i E'_i =0$. At least one the $q_i$'s is different from zero, so for concreteness let $q_M \neq 0$. We can then solve for $E'_M$ to get
\be
E'_M = q_{M}^{-1} \sum_{i=1}^{M-1} E'_i.
\ee
Define
\be
E^{(1)}_i := q_{M}^{-1} E'_i ,\quad i=1,\ldots,M-1.\label{E1i}
\ee
Then the electric fields $E'_i,i=1,\ldots,M$  can be expressed as integer linear combinations of the $M-1$ electric fields $E^{(1)}_i$ (\ref{E1i}). If $E^{(1)}_1,\ldots,E^{(1)}_{M-1}$ are algebraically independent, we are done. Otherwise we apply the above procedure to the $M-1$ electric fields $E^{(1)}_i$ to obtain a new set of $M-2$ electric fields in terms of which the rest are expressed as integer linear combinations. The procedure is iterated until one obtains an algebraically independent set. 

\subsection{Eq. (\ref{plfint})} \label{plfintapp}
By linearity of the PLF, it is enough to consider the special case where $a_l$ takes the form
\be
a_l(g_1, \ldots, g_n, u_1, \ldots, u_N) =\tilde{a}(g_1,\ldots,g_n) (u_1)^{m_1} \ldots (u_N)^{m_N} ,\label{ffactor}
\ee
where $m_I \in \integers, I=1,\ldots,N$.
In such case, we have
\ba
\w(a) &=& \bra 0,0 | \tilde{a}(\hat{h}_{e_1},\ldots,\hat{h}_{e_n})\cdot 0 , \sum_{I=1}^N m_I E_I  \ket \\
 &=& \bra 0| \tilde{a}(\hat{h}_{e_1},\ldots,\hat{h}_{e_n}) |0 \ket_{\text{LQG}} \delta_{0,\sum_{I=1}^N m_I E_I } \\
 &=& \int_{SU(2)^n}  \tilde{a}({g})d \mu  \prod_{I=1}^N\delta_{0,m_I}. \label{lastwf}
\ea
Here we used that standard rewriting of the LQG PLF in terms of $SU(2)$ integrals \cite{ttbook},  the  algebraic independence of the electric fields (\ref{algind}), and the basic inner product $\bra s_1, E_1 | s_2, E_2 \ket =\bra s_1 | s_2\ket_{\text{LQG}} \delta_{E_1 E_2}$.

As in the treatment of $\reals$-Bohr \cite{ttbook,vel2}, we notice that the last factor in (\ref{lastwf}) corresponds to an integral over $U(1)^N$ with normalized Haar measure:
\ba
\int_{U(1)^N} d \mu (u_1)^{m_1} \ldots (u_N)^{m_N}&= &\prod_{I=1}^{N} \int \frac{d \theta_I}{2 \pi}e^{i m_I \theta_I} \label{intu1} \\
&=&\prod_{I=1}^N\delta_{0,m_I}. 
\ea
Substituting the Kronecker deltas in (\ref{lastwf}) by (\ref{intu1}) we recover (\ref{plfint}) for the special case of $f$ given by (\ref{ffactor}). By linearity, it follows that (\ref{plfint}) holds for general algebra elements.

\section{Projective limit} \label{projlimapp}
In this appendix we give further details on the projective limit space and clarify the relation between the use of preordered and partially ordered label sets. To simplify the discussion, we first describe in detail the case of standard LQG in section \ref{plimapp1}. In section \ref{plimapp2} we give the partially ordered label set description of $\Abar$. In section \ref{plimapp3} we show cylindrical consistency of the Haar measures on $G_l$ and $G_{\hat{l}}.$
\subsection{Relation between  preorder and partially ordered label sets for holonomy probes.} \label{plimapp1}

In the usual construction, the label set is given by subgrupoids of $\P$ generated by finitely many edges. Let $\L_\lqg$ be such label set so that   $L \in \L_\lqg$ denotes a subgrupoid of $\P$ generated by a finite number of edges. The relation  $L' \geq L$ iff $L$ is a subgroupoid of $L'$, makes $\L_\lqg$ a partially ordered directed set. The compact space associated to $L \in \Lh$ is: 
\be
\A_L := \hom(L,SU(2)),
\ee
and the projections $p_{L L'}$ are defined by restriction:  $y_{L'} \in \A_{L'}$ induces a homomorphism on any subgroupoid $L \leq L'$ by simply restricting the action of $y_{L'}$ to $L$. Let us denote by $\Abar_\lqg$ the resulting projective limit space, as described in \cite{velhinho,ttbook}.

The corresponding ingredients in our construction are:  The label set $\Lh$, the compact spaces $G_\gamma= SU(2)^n$, and  the projections $p_{\gamma \gamma'}$  determined by the way edges in $\gamma$ are decomposed in terms of edges of $\gamma''$, see section \ref{charhba}. It is easy  to verify the compatibility of the projections with the relation  `$\geq$' in the sense described in section \ref{sec5}. The corresponding projective limit space can be constructed completely analogous to $\Abar_\lqg$: The `ambient' space $G_{\infty}:= \prod_{\gamma \in \Lh} G_\gamma$ with the  Tychonov topology (the weakest making the canonical projections to $G_\gamma$ continuous) is compact and Hausdorff \cite{ttbook}. The projective limit space  is the subset  $\Ah \subset G_\infty$ of points in $G_\infty$ satisfying  consistency conditions with the projections:
\be
\Ah: = \{ \{ x_\gamma \} \in G_\infty \; | \; p_{\gamma,\gamma'}(x_{\gamma'})=x_\gamma, \;  \forall \gamma' \geq \gamma \}  .
\ee
$\Ah$ is given the topology induced by $G_{\infty}$, and the same proof \cite{ttbook} that $\Abar_\lqg$ is closed goes through here as well\footnote{Lemma 6.2.10 in \cite{ttbook} can be repeated to show that every convergent net $\{x^\alpha_l \}$ in $G_{\infty}$ such that $\{x^\alpha_l \}$ is in $\Ah$ for any $\alpha$, converges to a point in $\Ah$ }. By the same arguments as for $\Abar_\lqg$, it follows that $\Ah$ is a compact, Hausdorff space.

Let us see that  $\Abar_\lqg$ and $\Ah$ are homeomorphic. A bijection between the two spaces can be  given as follows. Denote by  $\P_\gamma \in \L_\lqg$ the subgroupoid generated by $\gamma$ and let 
\be
\gen(L) =\{ \gamma \in \Lh \; | \; \P_\gamma= L \},
\ee
 be the set of all possible `generators' of a given $L \in \L_\lqg$. For   $\gamma,\gamma' \in \gen(L)$ we have that $\gamma' \geq \gamma$ and $\gamma \geq \gamma'$.   $p_{\gamma, \gamma'}$ then defines a homeomorphism between $G_{\gamma'}$ and $G_\gamma$ with inverse given by $p_{\gamma', \gamma}$.
  An element $y_L \in \A_L$ defines a point in $G_\gamma, \gamma \in \gen(L)$ by \cite{velhinho}:
\ba
\rho_\gamma : & \A_L &  \to  G_\gamma, \quad \gamma=(e_1, \ldots,e_n) \in \gen(L),\\
&y_L &  \mapsto (y_L(e_1),\ldots,y_L(e_n)) =:y_L(\gamma)  .
\ea
The points  $x_\gamma:= y_L(\gamma) \in G_\gamma$ for each $\gamma \in \gen(L)$  satisfy the consistency conditions
\be
p_{\gamma \gamma'}(x_{\gamma'})=x_\gamma , \quad  p_{\gamma' \gamma}(x_{\gamma})=x_{\gamma'} ,\quad \gamma,\gamma' \in \gen(L). \label{xggp}
\ee
Conversely,  any $x_\gamma \in G_\gamma$ determines a homomorphism in $ \A_{\P_\gamma} $ by
\ba
\sigma_\gamma : & G_\gamma &  \to  \A_{\P_\gamma} \label{defsigmagamma} \\
& x_\gamma &  \mapsto y_{\P_\gamma} \; | \; y_{\P_\gamma}(\gamma)=x_\gamma,
\ea
and given  $x_\gamma$ and $x_{\gamma'}$ satisfying   (\ref{xggp}), they define the same homomorphism. The continuous maps $\rho_\gamma$ and $\sigma_\gamma$ above are inverses of each other.   Let:
\ba
\tilde{p}_\gamma & := & \rho_\gamma \circ p_{L} : \Abar_\lqg \to G_\gamma \\
\tilde{p}_L & := & \sigma_\gamma \circ p_\gamma : \Ah \to \A_L , \quad \gamma \in \gen(L).
\ea
Then it is easy to verify that the maps
\ba
\rho  : &\Abar_\lqg \to \Ah, \quad & \rho(\{x_{L'} \}) :=  \{ \tilde{p}_\gamma(\{x_{L'}\}) \} \\
\sigma  : &\Ah \to \Abar_\lqg  , \quad  & \sigma(\{x_{\gamma'} \}): = \{ \tilde{p}_L(\{x_{\gamma'}\}) \},
\ea
are inverse of each other and that they provide a bijection  between  $\Abar_\lqg$ and $\Ah$. Finally this bijection is clearly a homeomorphism: The topology on $\Ah$ generated by the projections $p_\gamma$ coincides with that generated by the projections $\tilde{p}_L$ by virtue of continuity and invertibility of the maps $\rho_\gamma$ and $\sigma_\gamma$.

\subsection{Partially ordered directed label set for $\Abar$} \label{plimapp2}
\def\Lh{\hat{\mathcal{L}}}
\def\lh{\hat{l}}
We describe here the partially ordered directed set relevant for $\Abar$. 
Define an equivalence relation in $\L$ by $l \sim l'$ iff $l\geq l'$ and $l' \geq l$. Let $\Lh=\L/\sim$ be the corresponding quotient space.  We denote by  $\lh$ elements of $\Lh$.  On $\Lh$ we can define the relation $\lh \geq \lh'$ iff $l \geq \l'$ for some $l \in \lh$ and $l' \in \lh'$. It is easy to verify that this relation is well defined (independent on the choice of representatives $l$ and $l'$), and that it defines a partial order on $\Lh$, since by construction $\lh \geq \lh'$ and $\lh' \geq \lh$ implies $\lh = \lh'$. It is easy to verify that the directed set property of $\L$ implies that $\Lh$ is a partially ordered directed set. 

An intrinsic characterization of $\lh$ can be given as in the previous section:  $\lh  \approx (\P_\gamma,\E_\Upsilon) =:(\hat{\gamma},\hat{\Upsilon})$, where $(\gamma,\Upsilon) \in \lh$. The pair  $(\P_\gamma,\E_\Upsilon)$ is independent of the choice of representative, 
and the $\geq$ relation defined above for $\Lh$ corresponds to:  $\lh \geq \lh'$ iff the pair  of subgrupoids associated to $\lh'$ are subgroupoids of the pair associated to $\lh$.

We now describe the spaces and projections associated to the label set $\Lh$. Given $l,l' \in \lh$, we have maps $p_{l l'}: G_{l'} \to G_{l}$ and $p_{l' l}: G_{l} \to G_{l'}$. The consistency condition of the maps  imply that $p_{l l'} \circ p_{l' l}= \id_{G_l}$ and $p_{l' l} \circ p_{l l'}= \id_{G_{l'}}$, so that $p_{l l'}$ is a diffeomorphism between $G_{l'}$ and $G_{l}$. The space $G_{\lh}$ could then be defined as $G_l$ for some fixed representative $l \in \lh$. In the present case however the label sets have additional structure that allows for a more intrinsic definition of $G_{\lh}$. By the same argument as in the previous section, it is easy to verify that an element $g \in G_l$ defines a pair of homomorphisms $\hat{g}=(g^{\rm H},g^{\rm B}) \in \hom(\P_\gamma,SU(2)) \times \hom(\E_\Upsilon,U(1))$, with $l =(\gamma,\Upsilon)$. Further,   $p_{l' l}(g)\in G_{l'}$ defines the same pair of homomorphisms $\hat{g}$ for any  $l' \in \lh$. Thus we set $G_{\lh}:=\hom(\P_\gamma,SU(2)) \times \hom(\E_\Upsilon,U(1))$ with $(\gamma,\Upsilon) \in \lh$. The definition is independent of the choice of representative  $(\gamma,\Upsilon) \in \lh$. Finally, the projections $p_{\lh \lh'}$ can be defined, as in the previous section, by restriction of the homomorphisms to the corresponding subgroupoid. Such definition is then compatible with the projections $p_{ll'}$. 

The discussion of the previous section can be easily adapted to the present case to conclude that the projective limit space associated to $( \Lh,\{ G_{\lh} \} ,\{ p_{\lh \lh'} \})$ is homeomorphic to $\Abar$.
\subsection{Projective consistency of the Haar measures on $G_l$}\label{plimapp3}
Let $G_l = SU(2)^n \times U(1)^N$ where $n$ and $N$ are the number of independent edges and electric fields in $l$. Let $\mu_l$ be the normalized Haar measure on $G_l$ so that $\mu_l$ is a product of Haar measures on the $SU(2)$'s and $U(1)$'s factors.  We want to show that $(p_{l,l'})_* \mu_{l'}= \mu_l$ whenever $l' \geq l$ so that $\{ \mu_l , l \in \L \}$ define a consistent family of measures. Recall that the maps  $p_{l l'}$, described in section \ref{sec4C} are `block diagonal' i.e. do not mix $SU(2)$ factors with $U(1)$ ones. Given $(\gamma',\Upsilon') \geq (\gamma,\Upsilon)$, $p_{l l'}$ is determined by maps
\ba
g_i & = & p_i(g'_1,\ldots,g'_{n'}), \quad i=1,\ldots,n \\
u_I & = & P_I(u'_1,\ldots,u'_{N'}), \quad i=1,\ldots,N.
\ea
The consistency condition then translates into  two separate conditions,  $p_* \mu_{n'}=\mu_n$ and $P_* \mu_{N'}=\mu_N$ where $\mu_{n}$ is the Haar measure on $SU(2)^n$, $\mu_{N}$ that of $U(1)^N$ and similarly for the primed quantities. The proof that $p_* \mu_{n'}=\mu_n$ is the same as the one used in standard LQG to show the cylindrical consistency of the $SU(2)$ Haar measures, see \cite{velhinho,aajurekhoop}.  Such a proof is mainly group theoretical, and so it should not be difficult to adapt it  to the $U(1)$ factors. Below we present an alternative proof for the cylindrical consistency of the $U(1)$ measures. 

Let $C(U(1)^N)$ be  the space of continuous functions  on $U(1)^N$.  Recall that $C(U(1)^N)$ is a $C^*$  algebra with norm  $||f||:=\sup_{u \in U(1)^N} |f(u)|$, so that in particular it is a normed vector space. Define the linear functionals $\Gamma$ and $\Gamma'$ on this space by:
\ba
\Gamma(f) & := &\int_{U(1)^N}f(u_1,\ldots,u_N) d \mu_N \\
\Gamma'(f) & := & \int_{U(1)^{N'}} f(P_1(u'),\ldots,P_N(u')) d \mu_{N'}  \label{u1mucyl}
\ea
where 
\be
P_I(u')= \Pi_{J=1}^{N'} (u'_J)^{q^J_{I}},
\ee 
with $q^J_{I}$  the integers that determine how electric fields of $\Upsilon$ are written in terms of those in $\Upsilon'$, see   Eq. (\ref{Ptilde}).
Showing $P_* \mu_{N'}=\mu_N$ is then equivalent to showing $\Gamma(f) =\Gamma'(f) \; \forall f \in C(U(1)^N)$. First, we note that both $\Gamma,\Gamma'$ are bounded and hence continuous with respect to the topology of $C(U(1)^N)$: Clearly $| \Gamma(f)| \leq ||f||$. For $\Gamma'$ we have:
\be
|\Gamma'(f)| \leq \sup_{u' \in U(1)^{N'}} |f(P(u'))| = \sup_{u \in U(1)^N} |f(u)|= ||f||,
\ee
where the first equality is due to the fact that the map $P:U(1)^{N'} \to U(1)^N$ is surjective. Next, let $\pol(U(1)^N) \subset C(U(1)^N)$ be the set of functions given by finite linear combinations of elements of the form $\Pi_{I=1}^{N}(u_I)^{m_I}$ with $m_I \in \integers$. Since $\pol(U(1)^N)$ is a * sub algebra of $C(U(1)^N)$ and separates points, it follow by the Stone-Weierstrass theorem that $\pol(U(1)^N)$ is dense in $C(U(1)^N)$. By the bounded linear transformation theorem \cite{ttbook} it is then enough to show that $\Gamma$ and $\Gamma'$ agree on this dense subset of $C(U(1)^N)$. Finally,  by linearity we can focus attention on an element $\Pi_{I=1}^{N}(u_I)^{m_I}$.
One finds (see Eq. (\ref{intu1})):
\be
\Gamma(\Pi_{I=1}^{N}(u_I)^{m_I}) = \Pi_{I=1}^N\delta_{0,m_I}
\ee
and
\ba
\Gamma'(\Pi_{I=1}^{N}(u_I)^{m_I})  & = & \int_{U(1)^{N'}}  \Pi_{I=1}^{N}(\Pi_{J=1}^{N'} (u'_J)^{q^J_I})^{m_I} \\
&=& \Pi_{J=1}^{N'} \delta_{0,{\sum_{I=1}^{N}m_I q^J_I}} \\
&=& \Pi_{I=1}^N\delta_{0,m_I},
\ea
where we used $\int_{U(1)} u^{m}= \delta_{0,m}$ and the rational independence of the electric fields:
\be
m_I q^J_I =0 \; \forall J  \iff    m_I q^J_I E'_{J} =0  \iff  m_I E_I =0 \iff m_I =0 \; \forall I .
\ee

We thus have shown that the measures $\{ \mu_l \}$ represent a family of consistent measures on  $( \L,\{ G_{l} \} ,\{ p_{l l'} \})$. The measures $\mu_{\lh}$ on $G_{\lh}$ are defined by the push forward of maps $\sigma_{l}: G_l \to G_{\lh}$ defined analogously as  $\sigma_\gamma$ in Eq. (\ref{defsigmagamma}). The consistency of the  measures  $\{\mu_l \}$ immediately implies the consistency of the measures $\{\mu_{\lh}\}$.

\end{document}